\documentstyle[12pt,aasms4]{article}

\def\ga{\gtrsim}
\def\la{\lesssim}
\def\etal{{\it et al.}}
\def\tfm{two-fluid model}
\def\eg{{\it e.g.,\ }}
\begin{document}
\begin{flushright}
	\today
\end{flushright}
\title{ Analytic solution for nonlinear shock acceleration \\
 in the Bohm limit.}
\author{M. A. Malkov}
\affil{ Max-Planck-Institut f\"ur 
Kernphysik, D-69029, Heidelberg, Germany; \\
malkov@boris.mpi-hd.mpg.de} 

\begin{abstract}
The selfconsistent steady state solution for a strong shock, significantly 
modified by accelerated particles is obtained on the level of a kinetic 
description, assuming Bohm-type diffusion.  The original problem that is 
commonly formulated in terms of the diffusion-convection equation for the 
distribution function of energetic particles, coupled with the thermal 
plasma through the momentum flux continuity equation, is reduced to a 
nonlinear integral equation in one variable.  The solution of this equation 
provides selfconsistently both the particle spectrum and the structure of 
the hydrodynamic flow.  A critical system parameter governing the 
acceleration process is found to be \(\Lambda \equiv M^{-3/4}\Lambda_1 \), 
where \( \Lambda_1 =\eta p_1/mc \), with a suitably normalized 
injection rate \( \eta \), the Mach number \( M \gg 1 \), and the cut-off 
momentum \( p_1 \).

We are able to confirm in principle the often quoted hydrodynamic 
prediction of three different solutions.  We particularly focus on the most 
efficient of these solutions, in which almost all the energy of the flow is 
converted into a few energetic particles.  It was found that (i) for this 
efficient solution (or, equivalently, for multiple solutions) to exist, the 
parameter \( \zeta =\eta\sqrt{p_0 p_1}/mc \) must exceed a critical value 
$\zeta_{\rm cr} \sim 1$ (\(p_0 \) is some point in momentum space 
separating accelerated particles from the thermal plasma), and \( M \) must 
also be rather large (ii) somewhat surprisingly, there is also an upper 
limit to this parameter (iii)~the total shock compression ratio \( r \) 
increases with \( M \) and saturates at a level that scales as \( r \propto 
\Lambda_1 \) (iv) despite the fact that \( r \) can markedly exceed \( r=7 
\) (as for a purely thermal ultra-relativistic gas), the downstream 
power-law spectrum turns out to have the universal index \( q=3\onehalf \) 
over a broad momentum range.  This coincides formally with the test 
particle result for a shock of \( r=7 \) (v) completely smooth shock 
transitions do not appear in the steady state kinetic description.  A 
finite subshock always remains.  It is even very strong, \( r_{\rm s} 
\simeq 4 \) for \( \Lambda \ll 1\), and it can be reduced noticeably if \( 
\Lambda \ga 1 \).
\end{abstract}
\section{Introduction}
Strong astrophysical shocks are widely believed to be the sites where the 
cosmic rays (CRs) are born.  Although the test particle calculation of the 
CR spectrum is straightforward (Krimsky 1977; Axford, Leer \& Skadron 1977; 
Bell 1978; Blandford \& Ostriker 1978), the acceleration efficiency, in 
other words the fraction of incoming flow energy that is converted into 
CR-gas internal energy, cannot be obtained within this theory.  There are 
two major problems associated with diffusive shock acceleration.  First of 
all even in the simplest test particle theory a self-similar power-law 
solution, isotropic to lowest order, is possible only at sufficiently large 
momenta where the solution is independent of any momentum scale and depends 
only on the shock compression ratio.  Thus, some difficulty in this 
description occurs already at lower energies where the thermal and the bulk 
flow velocity may enter the solution.  Consequently, the amplitude of this 
power-law spectrum is virtually unknown.  It is therefore necessary to 
calculate, first of all, the injection rate, i.e.  the number of particles 
that feed the acceleration process.  Injection is believed to operate at 
the plasma subshock.  Surprisingly, until recently there were no attempts 
to attack the injection problem analytically.  At the same time this 
problem was studied numerically (\cite{el85}; \cite{quest}; \cite{sch92}; 
\cite{kj95}) and phenomenologically (\cite{lee82}; \cite{zwd}) in great 
detail and from different points of view.  Given the subshock strength and 
in terms of the parameterized distribution of thermal particles leaking 
into the upstream region from downstream, the injection rate has been 
calculated analytically by \cite{mv95}.  The distribution of these leaking 
(and/or directly from the shock front reflected) ions that is crucial for 
the injection theory is in turn a problem of its own in collisionless shock 
physics (\cite{sag64}, \cite{ken:sag}).  It almost certainly depends on 
shock parameters like the orientation of the magnetic field, the ratio of 
upstream thermal and magnetic pressures, the Mach number, etc.  One 
simplified formulation of this formidable task that also allows one to 
calculate this distribution in a closed form, suitable for injection, has 
been suggested by this author (\cite{M96}, M96 hereafter) for a parallel 
shock, where the direction of the shock normal and the magnetic field 
coincide.  For other types of shocks the question is unsolved.  It may be 
said then that the injection problem in general has become more a problem 
of the shock dissipation mechanism, rather than the problem of shock 
acceleration.  It is however also true as we shall see in the sequel that 
these two latter aspects of collisionless shock theory cannot be treated 
independently.

The second major problem, the impact of the accelerated high-energy 
particles on the acceleration process itself may be even more dramatic.  
While the injection rate can be calculated given the subshock conditions, 
even though subject to the above limitations, and the actual subshock 
strength can be determined afterwards (when the pressure of accelerated 
particles is finally obtained, see Appendix A), the calculation of the 
spectrum of dynamically important high-energy particles should be 
intrinsically nonlinear.  The reason is that the energetic particles can 
significantly modify the flow structure over a large spatial scale $\sim 
\kappa(p_1)/u_1 $, where $\kappa(p) $ is the momentum dependent diffusion 
coefficient (in the Bohm limit one has $\kappa \propto p$).  Here $p_1$ is 
the upper cut-off momentum and $u_1$ is the flow speed far upstream.  Hence 
particles with different energies ``see'' different shock compression.  
Moreover, they produce the gradient of the flow velocity by themselves, and 
therefore couple the length scale with the momentum scale.  This means that 
a momentum scale-free particle spectrum is no longer possible unless the 
velocity profile is also scale invariant.  Since in a strongly modified 
high Mach number shock the total compression ratio exceeds the conventional 
value of `four' markedly, the partial pressure of stationary accelerated 
particles becomes a nonintegrable function of momentum without an upper 
momentum cut-off.  Therefore, in reality the steady state acceleration 
along with the underlying flow structure will critically depend on the 
losses at the upper cut-off momentum $p_1$ or, more precisely, they will be 
determined by the balance between these losses and the injection around 
some slightly suprathermal \footnote{We regard particles downstream as 
suprathermal if they can be scattered upstream elastically.} momenta $p 
\sim p_0 \ga p_{\rm th}$ (see appendix A).  Moreover, these losses have the 
effect of increasing the total compression ratio, boosting CR production 
even further.  One sees that also the form of the spectrum obtained within 
the test particle theory is dubious since the backreaction of accelerated 
particles on the flow is significant.

In this paper we present a solution of the following problem.  Consider a 
strong, stationary, plane shock, propagating at the Mach number \( M \gg 1 
\).  Suprathermal particles are steadily injected at the subshock and are 
then partly accelerated up to the cut-off momentum \( p_1 \gg mc \).  To be 
determined (as functions of \( M \) and \( p_1 \)) are: (i) The spectrum of 
accelerated particles.  (ii) The flow profile across the shock \( u(x) \).  
The latter implies the total compression ratio and thus the acceleration 
efficiency, again as function of \( M \) and \( p_1 \).  Of particular 
interest are the question of the uniqueness of the solution, and the 
strength of the subshock.

Physically, the injection rate \( \eta \) does not belong to the input 
parameters which are merely \( M \) and \( p_1 \).  It can be calculated 
selfconsistently with the help of (MV95, M96, and \cite{mv97a}, (MV97)), 
given the subshock strength to be obtained below.  Nevertheless, under 
certain restrictions explained in Appendix A, injection can be treated 
independently of the large-scale shock modification studied in this paper.  
We will pursue this approach in what follows, considering \( \eta \) as 
another input parameter.

There is no necessity to stress that this paper is not the first one to 
attack the problem of nonlinear shock acceleration analytically.  Besides 
the well known two-fluid and three-fluid approximations introduced by 
Axford, Leer, \& Skadron (1977), Drury \& V\"olk (1981), and, including the 
scattering wave field, by \cite{McKV} on the one hand, and the perturbative 
kinetic studies performed by Blandford (1980) and Heavens (1983) on the 
other, there exist a few separable solutions based on rather special 
assumptions about the functional dependence of \( \kappa(p,x) \) (\eg 
Drury, Axford \& Summers 1982; \cite{webb85}).  An alternative treatment 
was suggested by Eichler (1979, 1984).  The key step of his approach 
consists in replacing the true solution of the diffusion-convection 
equation ad hoc by a Heaviside function which is coordinate 
independent up to some distance \( x_0(p) \) upstream and is zero beyond 
it.  Unfortunately, such a `weak solution' does not satisfy the convection 
diffusion equation which may be proven by substitution.  Physically there 
is indeed no other scale height upstream for the energetic particles than 
the diffusion length \( \kappa (p)/u \) which could justify the 
introduction of a step-like behavior of the particle distribution.  The 
width of the transition zone between the spectrum downstream and its far 
upstream (zero) limit increases with \( p \) exactly as does the length 
parameter \( x_0(p) \) where the spectrum supposedly vanishes abruptly in 
Eichler's model.  As it was pointed out by Eichler (1984), such a behaviour 
of the solution would be possible if \( \kappa(p,x) \) were extremely 
large for \( x \le x_0(p) \) and zero otherwise.  Therefore, this solution 
belongs also to the category of solutions that require very special form of 
diffusion coefficient \( \kappa \).  There is no physical reason, however, 
to ascribe such a property to the CR diffusivity.

Notwithstanding the above criticism, the approach developed by Eichler 
contains the fruitful idea of `algebraization' of the problem.  In other 
words, it offers a way to reduce the number of independent variables by the 
introduction of some functional link between the \( x \) and \( p \) 
variables through a special boundary \( x=x_0(p) \) in phase space.  In 
addition, the issue of interrelation between injection and losses and their 
critical influence on the shock formation and acceleration efficiency was 
first addressed by Eichler (1979, 1984).  In principle some form of 
algebraization seems to be necessary for a successful analytic treatment.  
Otherwise computer studies will remain the only possibility to explore this 
difficult problem.  In contrast to Eichler we derive an algebraization that 
is based on the powerful technique of integral transforms.

We shall focus predominantly on the so called efficient solution (Fig.1), 
in which almost all the shock energy is converted into CR-gas internal 
energy.  Its relation to the inefficient, test-particle solution is the 
subject of a companion paper (Paper II).  By an extension of the technique 
developed below, we present in Paper II a unified description of all three 
solutions along with the corresponding bifurcation analysis.

The outline of this paper is as follows.  In sec.2 we introduce the 
necessary equations and discuss some of the basic approximations used in 
the further analysis.  In sec.3 we find a formal solution for the particle 
distribution in the shock precursor, that depends on the unknown postshock 
spectrum and on the hydrodynamic velocity in the shock transition.  We 
introduce a spectral function that couples the thermal gas with the 
energetic particles and plays a central role in the analysis.  In sec.4 we 
derive equations for the spectral function and for the hydrodynamic flow, 
investigate analytic properties of the spectral function and solve the 
equation for it.  Sec.5 deals with the flow profile.  In Sec.6 the particle 
spectrum is restored.  We summarize and discuss the results in sec.7.
\section{Basic equations and assumptions}
Consider a strong CR-modified shock propagating in the positive \( x 
\)-direction in a cold medium with the gas kinetic pressure \( P_{\rm g1} 
\) and the mass density \( \rho_1 \).  In the shock frame of reference the 
steady mass flow profile is defined as follows 
\begin{equation}
	U(x)= \left\{ 
	\begin{array}{cc}
		-u(x), & x \ge 0  \\
		
		-u_2, & x < 0,
	\end{array} \right.
	\label{U(x)}
\end{equation}
where \( u_2 \) is the (constant) downstream mass velocity, \( u(0+)=u_0 
\), and \( u(x) \to u_1=const\), as \( x \to \infty \).  The isotropic part 
of the particle distribution at sufficiently high momenta, and for all \( x 
\), is governed by the well known diffusion-convection equation (Parker 
1965; Gleeson \& Axford 1967)
\begin{equation}
	\frac{\partial}{\partial x} \left( U g - \kappa(p) \frac{\partial 
g}{\partial x}\right) =\frac{1}{3} \frac{dU}{dx} p\frac{\partial 
g}{\partial p}
	\label{c:d:p}
\end{equation}
Here the number density of CRs is normalized to \( 4\pi gdp/p \), \( 
\kappa \) denotes the particle diffusion coefficient that is assumed to be 
a monotonically increasing function of momentum \( p \).  To be specific, 
we will assume that \( \kappa(p)=\kappa_0 p/p_0=\kappa_1 p/p_1 \).  For \( x 
< 0 \) the solution to eq.(\ref{c:d:p}) is taken to be \( U(x)=-u_2=const 
\) and \( g=g_0(p) \).  For \( x \ge 0 \ g(x,p)\) satisfies the equation
\begin{equation}
	\frac{\partial}{\partial x} \left( u g + \kappa(p) \frac{\partial 
g}{\partial x}\right) =\frac{1}{3} \frac{du}{dx} p\frac{\partial 
g}{\partial p}
	\label{c:d}
\end{equation}
which should be solved together with the conditions  of overall
mass and momentum conservation
\begin{eqnarray}
	\rho u & = & \rho_1 u_1,
	\label{c:e}  \\
	P_{\rm c}+P_{\rm inj}+\rho u^2 +P_{\rm 
	g1}\left(\frac{u_1}{u}\right)^\gamma & = & \rho_1 u_1^2 +P_{\rm g1}
	\label{m:c:p}
\end{eqnarray}
Here \( \rho(x) \) is the mass density, \( \rho_1 =\rho(\infty), \ \gamma 
\) is the specific heat ratio of the thermal plasma, \( P_{\rm c} \) is the 
CR pressure
\begin{equation}
	P_{\rm c}(x)= \frac{4\pi}{3} mc^2 \int_{p_0}^{p_1}\frac{p 
dp}{\sqrt{p^2+1}} g(p,x)
	\label{P_c}
\end{equation}
and no seed particles are present, i.e.  \( P_{\rm c}(\infty) =0\).  The 
particle momentum \( p \) is normalized to \( mc \).  We regard as CRs all 
the particles that occupy the region in momentum space between the 
injection momentum \( p_0 \) and the cut-off momentum \( p_1 \), and we 
assume that \( p_0 < 1 \ll p_1 \).  In Bernoulli's integral (\ref{m:c:p}) 
we have isolated the contribution of suprathermal particles \( P_{\rm inj} 
\) that do not belong to the thermal upstream gas and should rather be 
regarded as a low-energy part of the CRs.  These particles appear just 
upstream from the gaseous subshock as a product of thermal leakage from the 
downstream medium and have practically nothing to do with the adiabatically 
compressed upstream thermal plasma represented by the last term on the 
l.h.s.  in eq.(\ref{m:c:p}) (see Appendix A).  In view of the astrophysical 
significance of strong shocks, we focus on the high Mach number limit.  For 
simplicity we confine our consideration to the case in which not only \( M 
\gg 1 \) but also
\begin{equation}
	\nu \equiv \frac{2}{M^2}\left(\frac{u_1}{u_0}\right)^\gamma \ll 1
	\label{nu}
\end{equation}
where \( M^2 = \rho_1 u_1^2/\gamma P_{\rm g1} \). Thus eq.(\ref{m:c:p}) 
rewrites 
\begin{equation}
	u(x) +\frac{1}{\rho_1 u_1} \left(P_{\rm c}+P_{\rm inj}\right) = u_1
	\label{m:c}
\end{equation}
One remark should be made concerning the approximation in eq.(\ref{m:c}).  
Namely, we neglected the gas pressure by virtue of assumption (\ref{nu}), 
comparing it with the ram pressure \( \rho_1u_1^2 \).  If the shock is 
significantly modified by CRs, one should compare the gas pressure with \( 
\rho_1u_1^2-P_{\rm c} \) instead and the condition may become more 
stringent, \( \nu u_1/u_0 \ll 1 \).  However, this may potentially be 
important only close to the subshock where also the \(P_{\rm inj} \)-term 
may be important, and the problem can hardly be resolved without turning to 
the subshock internal structure itself.  We will specify the corresponding 
length scale and discuss the validity of eq.(\ref{m:c}) in Appendix C, 
after the flow structure is determined.  The last equation we need is that 
for the gaseous subshock, in which equation we neglect the energy losses 
from the gas due to injection:
\begin{equation}
	\frac{u_0}{u_2}=\frac{\gamma+1}{\gamma-1+\nu u_1/u_0}
	\label{c:r}
\end{equation}
It is important to note that the subshock can still be diminished 
significantly due to the CR pressure and at least formally can even be 
smeared out completely when \( \nu=2u_0/u_1 \), without violating our 
assumption (\ref{nu}).  To summarize, 
eqs.(\ref{c:d},\ref{c:e},\ref{m:c}), and (\ref{c:r}) form the basis for the 
further analysis.
\section{Approximate solution to the diffusion-convection equation}
The main purpose of this section is to gain enough information about 
the solution of the diffusion-convection equation (\ref{c:d}) without 
specifying the flow profile \( u(x) \).  To this end we will use the flow 
potential
\begin{equation}
	\Psi=\int_{0}^{x}u dx
	\label{f:p}
\end{equation} 
as a new independent spatial variable instead of \( x \).  But first, we 
transform eq.(\ref{c:d}) to the following integro-differential equation
\begin{eqnarray}
	g(x,p) & = & e^{-\Psi(x)/\kappa}\left[g_0(p)- \right.
	\label{} \nonumber \\
	& & \left.  \frac{1}{3\kappa}\int_{0}^{x}e^{\Psi(x^\prime )/\kappa} 
	dx^\prime p\frac{\partial}{\partial 
	p}\int_{x^\prime}^{\infty}\frac{du}{dx''}g(x'',p) dx'' \right]
	\label{c:d:i}
\end{eqnarray}
Once the solution of this equation is found in terms 
of the unknown downstream spectrum \( g_0(p) \), the latter can also be 
determined from the equation immediately following from eq.(\ref{c:d})
\begin{equation}
	\kappa \left. \frac{\partial g}{\partial x}\right|_{x=0+} = 
\frac{\Delta u}{3} \left(p \frac{\partial g_0}{\partial p}-3 g_0\right)
	\label{b:c}
\end{equation}
One sees that iterates \( g^{(n)} \) of eq.(\ref{c:d:i}), starting with 
\begin{equation}
	g^{(0)}=g_0(p)\exp\left(-\Psi(x)/\kappa\right) 
	\label{g:unm}
\end{equation}
yield a uniformly valid expansion in the case \( (u_1-u_0)/u_1 \ll 1\) 
(weak modification).  In fact, \( g^{(0)} \) is already a good 
approximation in this case.  Let us start our consideration of the strongly 
modified shock from the region \( \Psi/\kappa \ll 1 \).  Substituting \( 
g^{(0)} \) into the r.h.s.  of eq.(\ref{c:d:i}) we can generate a formal 
Neuman series in \( \Psi/\kappa \).  Indeed, calculating the internal 
integral by parts and retaining formally only the leading term in \( 
1/\kappa \), as a first iteration we get
\begin{equation}
	g^{(1)}=g_0 e^{-\Psi/\kappa}\left(1-\beta\frac{\Psi}{\kappa}\right)
	\label{g^1}
\end{equation}
where
\begin{equation}
	\beta \equiv -\frac{1}{3}\frac{\partial \ln g_0}{\partial \ln p}
	\label{bet}
\end{equation}
To the same order in \( \Psi/\kappa \) we can rewrite \( g^{(1)} \) in the 
same form as \( g^{(0)} \), but with a renormalized diffusion coefficient
\begin{equation}
	g^{(1)}=g_0(p)\exp \left\{-\frac{1+\beta}{\kappa}\Psi\right\}
	\label{ren:g}
\end{equation}
Since \( \beta \) may be rather small numerically, (\eg for the test 
particle solution with the compression ratio \(r=7\), one obtains \( 
\beta=1/6 \)) and because \( g^{(0)} \) provides a correct asymptotic 
behavior of the solution of eq.(\ref{c:d}) for \( \Psi \to \infty \, (du/dx 
\to \infty )\), one may conjecture that the solution in the form of 
eq.(\ref{ren:g}) is also good for larger values of \( \Psi/\kappa \).  That 
this is true, is demonstrated in Appendix B.  Therefore, we may take
\begin{equation}
	g=g_0(p)\exp \left\{-\frac{1+\beta}{\kappa}\Psi\right\}
	\label{c:d:sol}
\end{equation}
as a formal solution to the diffusion-convection equation.  The function \( 
g_0(p) \), or equivalently, \( \beta \) is yet to be obtained from 
eq.(\ref{b:c}) whereas the function \( \Psi(x) \) must come from the 
Bernoulli's integral (\ref{m:c}).
\subsection{Equation for the particle spectrum}
The solution (\ref{c:d:sol}) is valid for \( \Psi/\kappa_1 <\delta^{-1}\), 
where \( \delta \ll 1 \), and its region of validity has, in fact, also a 
lower bound given by eq.(\ref{l:ineq}) (see Appendix B).  As it will be 
shown later, this restriction concerns only a very small region at the 
origin for the solutions of interest.  Clearly, this may be important for 
\( \partial g/\partial \Psi \) at small \( \Psi \) but not for \( g \) 
itself since the latter is continuous at \( \Psi=0 \).  Thus, some care 
should be exercised in calculating the l.h.s.  of eq.(\ref{b:c}).  In fact, 
it is sufficient to rewrite this equation with the help of eq.(\ref{c:d:i}) 
as
\begin{equation}
	-\frac{\partial \ln g_0}{\partial \ln p} =\frac{3u_2}{\Delta 
u}+\frac{1}{\Delta u g_0}\frac{\partial}{\partial \ln p} 
\int_{0+}^{\infty} \frac{du}{dx} g(x,p) dx
	\label{eich:eq}
\end{equation}
This equation can also be derived directly from eq.(\ref{c:d}) by means of 
integration over \( x \) between \( 0- \) and \( +\infty \) (Eichler 1979).  
In contrast to eq.(\ref{b:c}) it does not contain the uncertain spatial 
derivative of the solution at the origin and we may substitute the solution 
(\ref{c:d:sol}) into eq.(\ref{eich:eq}) which yields
\begin{equation}
	 -\frac{1}{3} \frac{\partial \ln g_0}{\partial \ln p} = 
\frac{u_2+\frac{1}{3}\frac{\partial V}{\partial \ln p}}{\Delta u+ V}
	\label{g0}
\end{equation}
Here we have introduced the function \( V\) that will play a central role 
in our further analysis
\begin{equation}
	V(p)=\int_{0+}^{\infty}\frac{du}{dx} 
	dx\exp\left[-\frac{1+\beta}{\kappa}\Psi(x)\right]
	\label{V}
\end{equation}
This function reflects explicitly a degree of shock modification.  In an 
unmodified shock \( V(p) \equiv 0 \), since then \(du/dx=0 \) in the 
upstream region; the spectral index \( q=3\beta \) is just the conventional 
\( q = 3u_2/(u_1-u_2) \), (see eq.(\ref{g0})).  In general \( 0 \le V(p) 
\le u_1 -u_0 \) and \( V(p) \to u_1 -u_0 \) as \( p \to \infty \).  Even if 
the shock is appreciably modified, one may show (Appendix C) that at small 
\( p \ga p_0\)
\begin{equation}
	V(p) \ll \Delta u
	\label{V_0:est}
\end{equation}
and \( V \) may be neglected in eq.(\ref{g0}).  The spectral index then 
corresponds simply to the subshock compression ratio, and at lower momenta 
we have
 \begin{equation}
 	q \simeq q_0=\frac{3u_2}{u_0-u_2},\quad g_0(p)=Q_{\rm inj}
	\left(\frac{p}{p_0}\right)^{-q_0}
 	\label{low:m}
 \end{equation}
The injection solution (Appendix A) produces essentially the same 
asymptotic result for \( p \ga p_0 \), yielding thus the injection rate \( 
Q_{\rm inj} \).  The solution \( g_0(p) \) can be obtained then for all \( 
p \) using eq.(\ref{g0}).  To this end an independent equation for \( V(p) 
\) should be derived.  This will be done in the next section.
 \section{Integral equation for the spectral function}
In the previous section we have seen that the function \( V(p) \), that we 
term the spectral function, is directly related to the spectral slope 
through eq.(\ref{g0}).  If \( V(p) \) were defined as an integral starting 
at \( 0- \), the subshock jump would also be incorporated into the spectral 
function as a point set component of the measure \( u(x) \) in the integral 
(\ref{V}).  We will use this `full' spectral function occasionally, 
denoting it by \( \bar V(p) \equiv V(p)+\Delta u \).  We emphasize that in 
addition to the link with the particle spectrum, this function contains all 
the information about the hydrodynamic flow structure.  Indeed, introducing 
a new variable
 \begin{equation}
 	s=\frac{1+\beta(p)}{\kappa(p)}
 	\label{s}
 \end{equation}
 the function \( V \) can be recast as the Laplace-Stieltjes transform of 
 the flow velocity \( u(\Psi) \)
 \begin{equation}
   	V(s)=\int_{0+}^{\infty} e^{-s\Psi} du(\Psi), \quad {\rm and} \quad
	\bar V(s)=\int_{0-}^{\infty} e^{-s\Psi} du(\Psi)
   	\label{V:L}
   \end{equation}
that maps \( u(\Psi) \) onto \( V(s) \), or in other words, \( V \) 
provides the spectral representation of the flow \( u(\Psi) \) in terms of 
the particle momentum \( p \) by virtue of eqs.(\ref{s},\ref{V:L}).  Yet 
another interpretation of \( V(p) \) is that a particle with momentum \( p 
\) samples the flow compression \( \delta u= u(x)-u_0 \) upstream, where \( 
x \) is defined by the relation \( (1+\beta)\Psi(x)/\kappa(p) \sim 1 \).

After these short remarks concerning the usefulness of the spectral 
function \( V \) for describing the CR-gas coupling, we note that the only 
equation available to derive an equation for \( V \) is the Bernoulli's 
integral (\ref{m:c}).  Using eq.(\ref{c:d:sol}) the latter can be rewritten 
as
\begin{equation}
	\frac{d \Psi}{dx} + F(\Psi) = u_1
	\label{m:c:f}
\end{equation}
where
\begin{equation}
	F(\Psi) = \eta u_1 \int_{p_0}^{p_1} \frac{p \tilde g_0 dp }  
	{\sqrt{p^2+1}}\exp\left[-\frac{1+\beta(p)}{\kappa(p)}\Psi 
\right]+\frac{P_{\rm inj}}{\rho _1 u_1}
	\label{Fb}
\end{equation}
Here the injection rate is given by
\begin{equation}
	\eta=\frac{4\pi}{3}\frac{mc^2 Q_{\rm inj}}{\rho u_1^2} \simeq 
	\frac{q_0}{3}\frac{mn_{\rm c}c^2}{\rho_1 u_1^2}
	\label{eta}
\end{equation}
and \( n_{\rm c} \ll \rho_1 /m \) denotes the number density of the CRs.  
The low energy asymptotics of \( \tilde g_0 \) is now fixed by \( \tilde 
g_0 = (p_0/p)^{q_0}, \, p \ga p_0 \), i.e \( g_0(p)=Q_{\rm inj} \tilde 
g_0(p) \) (see eq.(\ref{low:m})).  We will omit the tilde sign in what 
follows.  The quantity \( \eta \) is regarded as a known function of the 
subshock parameters, that can be inferred from the injection theory MV95, 
M96, MV97.

For \( \Psi > 0 \) the pressure integral (\ref{Fb}) is cut by the exponent 
at small \( \kappa (p) \la \Psi \).  This means that starting from some 
\(\Psi \ga \kappa (p_0) \) the first term on the r.h.s.  of eq.(\ref{Fb}) 
becomes virtually independent of \( p_0 \) since, due to the normalization 
\( g_0(p_0)=1 \), the function \( g_0 \) depends also on \( p_0 \) in such 
a way that \( \eta g_0 \) is \( p_0 \)-invariant.  For smaller \( \Psi \) 
(\( 0 \le \Psi < \kappa (p_0) \)) only the sum of the two terms on the left 
hand side (l.h.s.) is independent of \( p_0 \) and the contribution \( 
P_{\rm inj} \) that has the same scale in \( \Psi \) might be important.  
This is, however, a very small region compared to the total precursor 
height \( \Psi_{\rm p} \sim \kappa (p_1) \gg \kappa (p_0) \), in the case 
of strongly modified shocks we are interested in.  Therefore we may ignore 
here this region and omit the term \( P_{\rm inj} \).

Taking the \( x  \) derivative of eq.(\ref{m:c:f}), multiplying the 
result by \( \exp(-s\Psi) \) and integrating with respect to \( x \) we get
 \begin{equation}
 	V(s)+\int_{0}^{\infty}F^{\prime}(\Psi) e^{-s\Psi} d\Psi =0
 	\label{V:F}
 \end{equation}
Returning to the momentum \( p \)  (eq.(\ref{s})), for efficient 
solutions from (\ref{Fb}) we obtain
\begin{equation}
 	V(p) - \eta u_1 \int_{p_0}^{p_1} \frac{p' dp'  
	g_0(p')}{\sqrt{p'^2+1}} \frac{s(p')}{s(p')+s(p)}=0
 	\label{}
 \end{equation}
(Note that for inefficient, weakly modified solutions, the function \( s(p) 
\) should be set to \( s=1/\kappa \) in accordance with eq.(\ref{g:unm}), 
instead of eq.(\ref{s}).) Resolving eq.(\ref{g0}) for \( g_0 \) and using 
the normalization \( g_0(p_0)=1 \) in place of the last equation we obtain 
the following equation for \( V \)
 \begin{equation}
  	V(p)=\eta u_1 \int_{p_0}^{p_1} \frac{p' dp' }{\sqrt{p'^2+1}} 
  	\frac{s(p')}{s(p')+s(p)} \frac{\Delta u +V(p_0)}{\Delta u +V(p')} \exp 
  	\left[-3 u_2\int_{p_0}^{p'} \frac{d \ln p''}{\Delta u+V(p'')}\right]
  	\label{i:eq:p}
  \end{equation}
where the function \( s  \) may be written as (eqs.(\ref{bet},\ref{g0}))
  \begin{equation}
  	s(p)=\frac{u_0+V(p)+\frac{1}{3} \frac{\partial V}{\partial \ln 
  	p}}{\kappa(p)\left(\Delta u+V(p)\right)}
  	\label{s2}
  \end{equation}
in the case of efficient solutions, and \( s=1/\kappa(p) \) for inefficient 
ones.  Equation (\ref{i:eq:p}) is a nonlinear integro-differential equation 
for the function \( V(p) \).  Mathematically, the advantage of this 
equation is that it is only an equation for one function of one variable.  
By contrast, the original equations (\ref{c:d},\ref{m:c}) form a 
quasi-linear system of the partial differential equation (\ref{c:d}) in two 
variables \( x,p \) under the algebraic condition (\ref{m:c}).  Moreover, 
provided that eq.(\ref{i:eq:p}) is solved, eqs.(\ref{m:c:f},\ref{V:F}) and 
(\ref{g0}) allow us to calculate both the flow profile \( u(x) \) and the 
particle distribution \( g(x,p) \).  Though eq.(\ref{i:eq:p}) is fairly 
complex, it can be easily solved numerically.  It might be tempting to do 
so.  However, such a shortcut would not allow much further insight into the 
physical nature of the solutions.  Therefore, in this paper, we shall 
pursue an analytic approach with the aid of a number of simplifying 
approximations.
\subsection{Simplification of the integral equation for the spectral 
function}
We first introduce a normalized spectral function
\begin{equation}
	J(p)= \frac{\Delta u+V(p)}{\Delta u+V_0} \equiv 
	\frac{\bar V(p)}{\Delta u ^*}
	\label{J:d}
\end{equation}
where \( V_0=V(p_0),\,  \Delta u^*=\Delta u + V_0 \), and substitute the 
relativistic Bohm diffusion coefficient
$$	\kappa=\kappa_0 p/p_0   $$
Using  a new variable \( t=\kappa_0 s(p) \) in place of \( p \), 
eq.(\ref{i:eq:p}) rewrites
\begin{equation}
	J(t)=A\int_{t_1}^{t_0}\frac{dt'}{t'+t}\frac{1}{t'J(t')}
	\exp\left[-\frac{3u_2}{\theta \Delta u^*}\int_{t'}^{t_0}
	\frac{dt''}{t''J(t'')}\right] +\frac{\Delta u}{\Delta u^*}
	\label{J}
\end{equation}
Here we have introduced the notations
	\begin{equation}
		 A=\eta p_0 \frac{u_1}{\Delta u^*}; 
 \quad t_0=\frac{u_0^*}{\Delta u^*}; \quad 
		u_0^*=u_0+V_0+\frac{1}{3}\left.\frac{\partial 
	V}{\partial \ln p}\right|_{p=p_0};
		\label{defs1}
	\end{equation}
	\begin{equation}
		 t_1=\frac{p_0}{p_1} 
	\left[V(p_1)+u_0 +\frac{1}{3}\left.\frac{\partial 
	V}{\partial \ln p}\right|_{p=p_1}\right]\left[V(p_1)+\Delta 
	u\right]^{-1} \equiv \frac{p_0}{p_1} \theta \ll 1  
		\label{defs}
	\end{equation}
Several comments should be made on eq.(\ref{J}).  Through an argument 
similar to that followed by eq.(\ref{V:F}) we have simplified in 
eq.(\ref{J}) the contribution of the nonrelativistic particles (\( p_0 \la 
p <1 \)) and used the Bohm diffusion coefficient \( \kappa \propto p \) for 
all \( p \) instead of a more general form of it, \eg \( \kappa \propto 
p^2/\sqrt{1+p^2} \).  We thus choose the simplest realistic form of the 
diffusion coefficient.  Clearly such a choice suggests logically to replace 
\( \sqrt{1+p^2} \to p \) in eq.(\ref{i:eq:p}).  The most serious potential 
problem that might be caused by such a simplified treatment is to overlook 
a relativistic peak at \( p \sim 1 \).  The situation here is quite similar 
to that considered in (\cite{mv96}), where the contribution of this peak to 
the CR pressure is shown to be negligible in comparison with that of the 
region \( 1 \ll p \le p_1 \) for the efficient solutions considered further 
in this paper.  These are mainly technical arguments for such a simplified 
treatment of eq.(\ref{V}).  The physical argument, as before, is an 
insignificant dynamical role of nonrelativistic particles.  Besides, we 
replaced \( ds/dp \) by
$$ \frac{ds}{dp} \simeq -\frac{\theta}{p} s(p) $$ 
which is strictly valid only for \( V(p) \gg u_0 \).  Except for the region 
\( p \ga p_0 \), this is precisely the property of efficient solutions, as 
we shall see.

Our next simplification of eq.(\ref{J}) is based on some assumption about 
its solution and this assumption will be justified later.  Since for 
efficient solutions \( J(t_1) \gg J(t_0)=1 \), \( J(t) \) is large in the 
dynamically most important part of the interval \( (t_1,t_0) \), i.e.  for 
\( t \ll t_0 \) or, equivalently, \( p \gg p_0 \).  If we assume that the 
estimate \( J > C t^{-\lambda}\) holds, with some positive constants \( C 
\) and \( \lambda \), then \( t' \) in the exponent of eq.(\ref{J}) can be 
replaced by \( t'=t_1 \), since the integral will be dominated by the upper 
limit \( t_0\).  Thus eq.(\ref{J}) can be written as
\begin{equation}
	J(t)=A^*\int_{t_1}^{t_0}\frac{dt'}{t'+t}\frac{1}{t'J(t')}+
	\frac{\Delta u}{\Delta u^*}
	\label{J:i}
\end{equation}
where we introduced the constant \( A^* \), to be determined from the 
following solvability condition
\begin{equation}
	A^*=A\exp\left[-\frac{3u_2}{\theta \Delta u^*}
	\int_{t_1}^{t_0}\frac{dt}{tJ(t)}\right]
	\label{A*}
\end{equation}
It will be specified after the solution to eq.(\ref{J:i}) is found. Now, it 
is convenient to rescale variables in eq.(\ref{J:i}) as follows
\begin{equation}
	\tau=\frac{t}{\sqrt{t_0 t_1}}, \quad 
	F=\sqrt{\frac{\varepsilon t_0}{A^*}} J
	\label{tau:F}
\end{equation}
where 
$$ \varepsilon =\sqrt{t_1/t_0} \ll 1$$
and then eq.(\ref{J:i}) rewrites as
\begin{equation}
	F(\tau)=\int_{\varepsilon}^{1/\varepsilon}\frac{d\tau'}{\tau'+\tau}
	\frac{1}{\tau' F(\tau')} + \frac{\Delta u}{\Delta u^*} 
	\sqrt{\frac{\varepsilon t_0}{A^*}}
	\label{F}
\end{equation}
The goal of the next two subsections is to find an asymptotic (\( 
\varepsilon \ll 1 \)) solution of this equation.
\subsection{Analytic properties of the spectral function}
The kernel of eq.(\ref{F}) belongs formally to the Carleman type (see, \eg 
\cite{mus}) and the theory of linear equations of this type is 
very well developed.  The problem is that eq.(\ref{F}) is nonlinear and it 
is also well known that the solutions in this case can be fundamentally 
different from those in the linear case.  In particular, multiple solutions 
and bifurcations may occur.  A simple iterative approach would be quite 
reasonable for seeking inefficient solutions of eq.(\ref{F}), or better, of 
the more general equation (\ref{i:eq:p}).  \footnote{It should be also 
mentioned that after some transformations, eq.(\ref{F}) can be reduced to a 
form resembling a well known equation in radiative transfer theory, the so 
called Ambartsumyan equation which is also solved normally by iterations 
(\cite{amb}).  The analogy is, however, not complete and the solutions are 
different.} In contrast to inefficient solutions, the efficient solutions 
being strongly nonlinear, cannot be obtained perturbatively.  Rather we 
will find the solution to eq.(\ref{F}) which converges to the exact one as 
\( \varepsilon \to 0 \).  Indeed, once \( \varepsilon \) is set to zero, 
eq.(\ref{F}) contains no parameters at all and should be solved exactly.  
Before we find this asymptotic (\( \varepsilon \ll 1 \)) solution to 
eq.(\ref{F}), it is useful to establish some of its simple properties.

Consider the following function in the complex \( \tau \)- plane.
\begin{equation}
	G(\tau)=\int_{\varepsilon}^{1/\varepsilon}\frac{d 
\tau'}{\tau'+\tau}\frac{1}{\tau' F(\tau')}
	\label{G}
\end{equation}
Under obvious integrability conditions for the unknown function \( F \) 
(see \eg \cite{mus}), the function \( G \) is a holomorphic function in the 
whole \( \tau \)- plane, cut along the line \( (-1/\varepsilon,-\varepsilon 
)\) and has the degree -1 at infinity (\( G \sim 1/\tau,\, \tau \to \infty 
\)).  This function is discontinuous across the cut and the jump is equal 
to \( \Delta F= -2\pi i/\tau F(-\tau) \).  Using the Plemelji formulae from 
eq.(\ref{G}) we thus have
\begin{equation}
	G(-\tau \pm i0)={\mathcal{P}} 
	\int_{\varepsilon}^{1/\varepsilon}\frac{d\tau'}{\tau'-\tau}
	\frac{1}{\tau'F(\tau')} \mp \frac{i\pi}{\tau F(\tau)},
	\label{G:Pl}
\end{equation}
where, as indicated, the integral should be taken as a Cauchy principal 
integral.  From eq.(\ref{F}) we infer that the function \( F- (\Delta 
u/\Delta u^*) \sqrt{\varepsilon t_0/A^*}\) possesses the same analytic 
properties as \( G \).  Thus, the solution \( F \) to be found from 
eq.(\ref{F}) has two branch points at \( -1/\varepsilon \) and \( 
-\varepsilon \) and the jump \( \Delta F \) across \( (-1/\varepsilon, 
-\varepsilon \)).
\subsection{Solution for the spectral function.}     
As we emphasized, the only small parameter available in eq.(\ref{F}) is \( 
\varepsilon \ll 1 \).  This is a very small parameter indeed! For shock 
acceleration in situations of the astrophysical interest the magnitude of 
\( 1/\varepsilon \sim \sqrt{p_1/p_0} \) may be as large as \( 10^4 \) (for 
typical SNR conditions).  For a galactic wind termination shock this 
parameter may be even several orders of magnitude larger (see \cite{jm85}).  
It is therefore tempting to set \( \varepsilon =0 \) in eq.(\ref{F}).  In 
singularly perturbed problems, however, it is good practice to keep 
singular points at their `exact' positions in all orders of approximation.  
The singular character of the perturbation in the parameter \( \varepsilon 
\) is quite obvious since the case \( \varepsilon=0 \) corresponds to the 
absence of a cut-off momentum (\( p_1=\infty \)), and the solutions are 
structurally different from those for \( \varepsilon \neq 0\).  Such 
physically critical quantities as \( P_{\rm c } \) and the precursor length 
may diverge as \( \varepsilon \to 0 \).  We therefore keep \( \varepsilon 
\) small but fixed at some critical points in our further analysis.

First, we represent eq.(\ref{F}) in the following equivalent form
  \begin{eqnarray}
 	 F(\tau)& = & \int_{-\varepsilon}^{\infty}\frac{d\tau'}{\tau'+\tau}
	\frac{1}{\tau' F(\tau')}+
 	\label{} \nonumber  \\
 	 & + & \frac{\Delta u}{\Delta u^*}\sqrt{\frac{\varepsilon 
	t_0}{A^*}}-\left(\int_{-\varepsilon}^{\varepsilon}+
	\int_{1/\varepsilon}^{\infty}\right) \frac{d\tau'}
	{\tau'+\tau}\frac{1}{\tau' F(\tau')}
 	\label{F1}
 \end{eqnarray}
Shifting then the lower singular point to the origin \( (\tau+\varepsilon=y 
\)) we rewrite the last equation as follows
\begin{equation}
	F(y)= \int_{0}^{\infty}\frac{dy'}{y'+y}\frac{1}{y'F(y')} 
+\sqrt{\varepsilon} R(y,F,\varepsilon)
	\label{F2}
\end{equation}
where
 \begin{eqnarray}
 	R & = & 
\frac{\Delta u}{\Delta u^*}\sqrt{\frac{t_0}{A^*}}+
\frac{1}{\sqrt{\varepsilon}}\int_{2\varepsilon}^{1/\varepsilon +\varepsilon}
\frac{dy'}{F(y')}\frac{1}{y'+y-2\varepsilon}\frac{1}{y'-\varepsilon} - 
\label{} \nonumber  \\
 	 &  & \frac{1}{\sqrt{\varepsilon}}\int_{0}^{\infty}
	\frac{dy'}{y'F(y')}\frac{1}{y'+y}
 	\label{R}
 \end{eqnarray}
After the ansatz
 \begin{equation}
 	F(y)=F_0(y)+\sqrt{\varepsilon}F_1 +...
 	\label{F:ser}
 \end{equation}
for \( F_0 \) we have
 \begin{equation}
 	F_0(y)=\int_{0}^{\infty}\frac{dy'}{y'+y}\frac{1}{y'F_0(y')}
 	\label{F_0}
 \end{equation}
Taking \( \Re y \) negative from the last equation we obtain
 \begin{equation}
 	F_0(-y \pm i0)= 
	{\mathcal{P}}\int_{0}^{\infty}\frac{dy'}{y'-y}\frac{1}{y'F_0(y')} 
	\mp \frac{i\pi}{yF_0(y)}
 	\label{}
 \end{equation}
Separating real and imaginary parts we get
\begin{eqnarray}
	 \Re F_0(-y) & = & {\mathcal{P}}\int_{0}^{\infty}
	\frac{dy'}{y'-y}\frac{1}{y'F_0(y')}
	\label{re:f} \\
	\Im F_0(-y \pm i0) & = & \mp \frac{\pi}{y F_0(y)} 
	\label{im:f}
\end{eqnarray}
It is seen that the function
 \begin{equation}
 	F_0(y)=\sqrt{\frac{\pi}{y}}
 	\label{F_0:sol}
 \end{equation}
satisfies both last equations and, therefore, eq.(\ref{F_0}) exactly.

Returning to the variable \( \tau \) we have \( 
F_0(\tau)=\sqrt{\pi/(\tau+\varepsilon}) \).  This solution is not uniformly 
valid and deviates from the true solution at large \( \tau \sim \varepsilon 
^{-1}\), i.e.  the formal validity range is \( \left|\tau\right| \ll 
1/\varepsilon \).  The origin of this nonuniformity is obviously in the 
transformation from eq.(\ref{F}) to eq.(\ref{F_0}) in which only the lower 
singular point \( \tau =-\varepsilon \) was kept at its correct position in 
accordance with the analytic properties, considered in the preceding 
subsection.  The second singular point, namely \( \tau =-1/\varepsilon \) 
has been shifted to infinity.  If \( y \sim 1/\varepsilon \) then the term 
\( \sqrt{\varepsilon } R \) in eq.(\ref{F2}) must also be taken into 
account.  The standard way to get a uniformly valid expansion is to treat 
the solution at large \( \tau \sim 1/\varepsilon \) in exactly the same way 
as we did at the left end of the interval \( \varepsilon,1/\varepsilon \) 
and to construct then a composite expansion.  This requires, generally 
speaking, the calculation of higher order terms of the expansion, i.e.  \( 
F_1 \) etc., since \( F_0 \) might become so small that \( F_0 \sim 
\sqrt{\varepsilon} F_1 \) (see eq.(\ref{F:ser})).  Our knowledge of the 
analytic properties of the solution reduces computations dramatically.  
Since we need just a uniformly valid zero order approximation, we may omit 
the second term in eq.(\ref{F}) and obtain the following equation
\begin{equation}
	\bar 
	F(\tau)=\int_{\varepsilon}^{1/\varepsilon}\frac{d\tau'}{\tau'+\tau}
	\frac{1}{\tau'\bar F(\tau')}
	\label{barF}
\end{equation}
Here \( \bar F \) denotes the uniformly valid zeroth order approximation 
to \( F \) in eq.(\ref{F}).  Assuming \( \tau \ll 1/\varepsilon \), from 
the above results we can write
\begin{equation}
	\bar F \simeq \sqrt{\frac{\pi}{\tau+\epsilon}}
	\label{barF:sol}
\end{equation}
Next we observe that the transformation \( \tau \mapsto 1/\tau, \, \bar F 
\mapsto \tau \bar F \) maps eq.(\ref{barF}) onto itself. To be specific we
transform \( (\tau,\bar F) \mapsto (\xi, H) \) as follows
\begin{equation}
	\xi=1/\tau, \quad H(\xi)=\bar F/\xi
	\label{}
\end{equation}
and for \( H \) we obtain
\begin{equation}
	H(\xi)=\int_{\varepsilon}^{1/\varepsilon}\frac{d\xi'}
	{\xi'+\xi}\frac{1}{\xi'H(\xi')}
	\label{H}
\end{equation}
According to eq.(\ref{barF:sol}) from the last equation we infer
$$H(\xi) \simeq \sqrt{\frac{\pi}{\xi+\varepsilon}} $$
which is a good approximation for \( \left|\xi\right| \ll 1/\varepsilon \). 
Returning to \( (\bar F, \tau) \)  and taking eq.( \ref{barF:sol}) into 
account we obtain the uniformly valid 
asymptotic result
\begin{equation}
	\bar F=\sqrt{\frac{\pi}{(\tau+\varepsilon)(1+\varepsilon \tau)}}.
	\label{barF:un}
\end{equation}
It may be seen that this method of uniformization  of the 
asymptotic solution of eq.(\ref{F}) works very well in the interval\( 
(-1/\varepsilon, -\varepsilon) \) and, what is important, restores the 
correct position of the branch point \( -1/\varepsilon \) that had 
``escaped'' to infinity in the asymptotic result (\ref{barF:sol}).  Besides 
that, the solution (\ref{barF:un}) exhibits a proper behavior at infinity 
according to the analytic properties discussed in the previous subsection.  
This method, however, might be still insufficient in the region \( \tau 
\sim 1/\varepsilon \), especially if \( t_0/A^* \) is large, and if then 
the \( \sqrt{\varepsilon} \) term on the r.h.s.  of eq.(\ref{F2}) must be 
taken into account.  On the other hand, the region \( \tau \sim + 
1/\varepsilon \) corresponds to the low momenta \( p \sim p_0 \) and in the 
efficient solutions particles from this part of phase space play no 
dynamical role.  We will not consider this region here.
\section{Structure of the shock transition}
The solution for the spectral function \( V(p) \) obtained in the preceding 
section contains the undetermined constant \( A^* \) that appears in the 
integral equation (\ref{J:i}).  This constant is in fact an eigenvalue of 
the nonlinear (singular) equation (\ref{J})\footnote{The fact that 
eq.(\ref{J}) possesses an `inhomogeneous term' on the r.h.s.  is irrelevant 
in the sense that the equation is nonlinear and that Fredholm's alternative 
cannot be applied.}.  Hence, the solution for \( V(p) \) found in the last 
section exists only if eq.(\ref{A*}) possesses real solutions for \( A^* 
\)\footnote{This is irrelevant for the inefficient solution, that can be 
directly obtained from eq.(\ref{J}) as a Neuman series in \( A \) provided 
that this parameter is sufficiently small.}.  Using the solution 
(\ref{barF:un}) and the notations (\ref{tau:F}) we rewrite the equation for 
\( A^* \) as follows
\begin{equation}
	A^*=A \exp\left[-\frac{3u_2}{\theta \Delta u^*}
	\sqrt{\frac{\varepsilon t_0}{A^*}}I\right]
	\label{A*1}
\end{equation}
where
\begin{equation}
	I=\int_{\varepsilon}^{1/\varepsilon}\frac{d 
\tau}{\tau F(\tau)}
	\label{I}
\end{equation}
This integral can be calculated conveniently by transforming it to an 
integral along the cut (eq.(\ref{im:f})) and then substituting \( F \) 
from the approximate solution (\ref{barF:un}). This procedure yields
\( I=\sqrt{\pi/\varepsilon} \). On the other hand the same integral can 
be calculated `exactly' by multiplying eq.(\ref{barF}) by \( 1/F \) and 
integrating both sides between \( \varepsilon \) and \( 1/\varepsilon 
\). The result reads
\begin{equation}
	\frac{1}{\varepsilon}-\varepsilon = \frac{1}{2}I^2
	\label{}
\end{equation}
or \( I \simeq \sqrt{2/\varepsilon} \).  This insignificant discrepancy is 
de facto explained in the last paragraph of the preceding section.  Namely, 
the corrections to the asymptotic result (\ref{barF:un}) in the region \( 
\varepsilon < -\tau < 1/\varepsilon \) are of the order of \( 
\sqrt{\varepsilon} \ll 1 \).  On the other hand, being integrated over an 
interval of the length \( 1/\varepsilon \), this correction produces the 
contribution \( 1/\sqrt{\varepsilon} \).  This correction does not change 
the asymptotic behavior of the solution (\ref{barF:un}) in the dynamically 
most important part \( \tau \sim \varepsilon \).  Substituting \( I = 
\sqrt{2/\varepsilon} \) into eq.(\ref{A*1}) we obtain
\begin{equation}
	A^*=\eta p_0 \frac{u_1}{\Delta u^*} \exp \left[-\frac{\Omega}
	{\sqrt{A^*}}\right]
	\label{A*2}
\end{equation}
where 
\begin{equation}
	\Omega = 
\frac{3}{\theta}\sqrt{2}
\frac{u_2\sqrt{u_0^*}}{\Delta u^{*3/2}}
	\label{omega}
\end{equation}
Clearly, eq.(\ref{A*2}) not always has solutions and we need the shock and 
subshock compression ratios \( r \equiv u_1/u_2 \) and \( r_{\rm s} \equiv 
u_0/u_2 \) to solve this equation for \( A^* \).
\subsection{General equation for the shock structure. 
The problem of uniqueness}
From eqs.(\ref{tau:F}) and (\ref{barF:un}) we have
\begin{equation}
	J(s) = \sqrt{\frac{\pi t_0 A^*}{\left(\kappa_0 s 
+t_1\right)\left(\kappa_0 s+t_0\right)}}
	\label{J(s)}
\end{equation}
where \( \kappa_0 s=t \).
Using the relation (\ref{J:d}) and inverting the integral in 
eq.(\ref{V:F}) for \( F^\prime (\Psi) \) we obtain
\begin{equation}
	F^\prime (\Psi) = -\frac{\Delta u^*}{2\pi 
i}\int_{-i\infty}^{+i\infty}e^{s\Psi} J(s) ds
	\label{F'}
\end{equation}
The last integral can be transformed to the following integral along the 
cut (see sec.4.2)
\begin{equation}
	F^\prime(\Psi)=\frac{\Delta u^*}{2\pi 
	i}\int_{-t_0/\kappa_0}^{-t_1/\kappa_0}
	e^{s\Psi}\left[J(s+i0)-J(s-i0)\right] ds
	\label{}
\end{equation} 
Substituting \( J(s) \) we deduce
\begin{equation}
	F^\prime(\Psi)=-\sigma e^{-a_+ \Psi} I_0(a_- \Psi)
	\label{}
\end{equation}
where
\begin{equation}
	\sigma =\frac{\Delta u^*}{\kappa_0}\sqrt{\pi t_0 A^*}, \quad a_\pm =
	\frac{1}{2 \kappa_0}(t_0\pm t_1)
	\label{}
\end{equation}
and \( I_0 \) denotes the modified Bessel function.  Using the boundary 
condition \( d \Psi /dx=u_0, \, \Psi =0+ \), from eq.(\ref{m:c:f}) we now 
obtain the following differential equation for the flow potential \( \Psi 
\)
\begin{equation}
	\frac{d 
	\Psi}{dx}=u_0+\sigma\int_{0}^{\Psi}e^{-a_+\Psi^\prime}
	I_0(a_-\Psi^\prime) d\Psi^\prime
	\label{fl:pot}
\end{equation}
which can be always integrated in closed form.  We shall describe the shock 
structure in the next subsection in more detail.  Here we determine the 
global characteristics of the flow.  Using the boundary condition \( 
d\Psi/dx \to u_1 \) as \( \Psi \to \infty \) and calculating the integral 
\footnote{It is worthwhile to note that this integral diverges as \( 
\sqrt{\Psi} \) if \( \varepsilon =0 \), i.e.  \(t_1=0 \) emphasizing the 
singular character of the \( \varepsilon \)- perturbation.} we arrive at 
the following relation for the compression ratios \footnote{Again, it is in 
principle possible to calculate here the integral without using the 
approximate solution (\ref{J(s)}) by the method applied for calculating the 
integral \( I \) cf.  eq.(\ref{I}).  The result would also differ by the 
factor \( \sqrt{2/\pi} \).  We ignore this small difference and use the 
eq.(\ref{fl:pot}) in what follows since it exactly conforms to the flow 
profile.}
\begin{equation}
	r = r_{\rm s}+(r_{\rm s}^*-1)\sqrt{\frac{\pi A^*}{t_1}}
	\label{r}
\end{equation}
where \( r_{\rm s}^*=r_{\rm s}+V_0/u_2 \simeq r_{\rm s} \equiv u_0/u_2 \).  
It should be emphasized that we have neglected here \( V_0 = V(p_0) \) in 
comparison with \(u_0 \), not because our solution \( V(p) \) has such a 
property-- it may be inaccurate at \( p\simeq p_0 \), see the last 
paragraph in the preceding section-- but rather by relying on an estimate 
that is made in Appendix C, eq.\ref{v:ov:u}.  Eq.(\ref{r}) should be solved 
together with eq.(\ref{A*2}) and with the subshock R-H relation 
(\ref{c:r}). If the subshock is significantly diminished, we note 
that this implies automatically \( r \gg r_{\rm s} \), because of our 
constraint (\ref{nu}) on the Mach number \( \nu \ll 1 \).  It should be 
emphasized that the condition \( r \gg r_{\rm s} \) is a characteristic of 
the strong CR shock modification rather than a decrease of \( r_{\rm s} \) 
as it is often asserted in the literature.  As we will see, a very strong 
shock modification may occur in the high Mach number limit also despite of 
\( r_{\rm s } \simeq 4 \).  From eqs.(\ref{A*2}) and (\ref{r}) we then 
obtain
\begin{equation}
	\sqrt{A^*}= \eta \sqrt{\pi p_0 p_1/\theta} \exp\left(-\frac{\Omega}
	{\sqrt{A^*}}\right)
	\label{A*3}
\end{equation}
The numerical factor \( \theta \) here was introduced earlier in eq.( 
\ref{defs}).  It will be determined after the particle spectrum has been 
found.

Denote \( \Omega/\sqrt{A^*}=z \).  We then have the following bifurcation 
equation for \( z \)
\begin{equation}
	 \Gamma ze^{-z}=1
	\label{bif:eq}
\end{equation}
where \( \Gamma =\eta \sqrt{\pi p_0 p_1\theta^{-1}}/\Omega \).  
Eq.(\ref{bif:eq}) has either two solutions, (if \( \Gamma >e \)) or no 
solution at all (\( \Gamma < e \), see, however, below).  At the 
bifurcation point \( \Gamma =e \) there exists a double root.  Therefore, 
these two solutions exist only if
\begin{equation}
	\eta\sqrt{p_0 p_1} 
>B \frac{\sqrt{r_{\rm s}}}{\left(r_{\rm s}-1\right)^{3/2}}
	\label{crit}
\end{equation}
where 
\begin{equation}
B=	3e \sqrt{\frac{2}{\pi \theta}}
	\label{B}
\end{equation}
We implied again that \(V_0 \ll \Delta u \) 
which will be shown to hold unless the subshock is very weak. In the case 
\( \Gamma -e \ll 1 \) this pair of solutions may be written as follows
\begin{equation}
	\sqrt{A^*_{\pm}} = \frac{\Omega}{1 \pm \sqrt{2(1-e/\Gamma)}} \nonumber
	\label{A:pm}
\end{equation}
Since \( \eta \) is bounded as \( r_{\rm s} \to 1\) (MV95), a very strong 
subshock reduction is prohibited by the criterion (\ref{crit}), and, 
therefore, a finite subshock must remain.  We shall return to the issue of 
subshock reduction later, after the flow profile is obtained.

This was the calculation at criticality, i.e.  where \( \Gamma \simeq e\) 
which occurs, say, at certain Mach number \( M \), provided that \( p_1 \) 
is fixed.  Consider now an acceleration regime that can be characterized by 
the condition \( \Gamma \gg 1 \), i.e.  far from the bifurcation point.  We 
then have from eq.(\ref{bif:eq}) for the first solution (\( z \ll 1 \))
\begin{equation}
	\sqrt{A^*} \simeq  \eta \sqrt{\pi p_0 p_1/\theta}
	\label{A1}
\end{equation}
and 
\begin{equation}
	\sqrt{A^*} \simeq \frac{\Omega}{\ln \Gamma}
	\label{A2}
\end{equation}
for the second one (\( z \gg 1 \)).  For the sake of convenience we call 
the solution (\ref{A1},\ref{A:pm}-) ``efficient'' and the solution 
(\ref{A2},\ref{A:pm}+) ``intermediate''.  As we argued earlier there exists 
also the third one, i.e.  a conventional test particle or ``inefficient'' 
solution of eq.(\ref{J}) at least for sufficiently small \( \eta \).  This 
solution cannot be obtained by means of the approach developed in Sec.4.3 
but it can be found directly from eq.(\ref{J}) as a Neuman series for \( A 
\ll 1 \).  The relations between these three solutions will be considered 
in Paper II.  In the case of a weak influence of the CRs on the flow the 
inefficient solution was studied perturbatively by Blandford (1980) and 
Heavens (1983) on the basis of the diffusion-convection equation 
(\ref{c:d}).  Here we concentrate on the efficient solution for which from 
eqs.(\ref{r}) and (\ref{A1}) we deduce
\begin{equation}
	r \simeq r_{\rm s} +\pi (r_{\rm s}-1) \eta p_1/\theta
	\label{r1}
\end{equation}
Note that the parameter \( p_0 \) which is irrelevant for the efficient 
solution has indeed disappeared from the result, as discussed in Appendix 
A.  The total compression ratio and, therefore, the acceleration efficiency 
depend only on the subshock compression ratio, on the injection rate \( 
\eta(r_{\rm s}) \), and on the upper cut-off momentum \( p_1 \).  Combining 
eq.(\ref{r1}) with the subshock R-H relation (\ref{c:r}) we obtain an 
equation for the single variable \( w \), given by
\begin{equation}
	w=\frac{r}{r_{\rm s}}M^{-\frac{2}{\gamma+1}}
	\label{w:def}
\end{equation}
The equation reads
 \begin{equation}
  	w=\mu+\lambda\left(1-w^{\gamma+1}\right),
  	\label{w:eq}
  \end{equation}
where
 \begin{equation}
 	\mu =M^{-\frac{2}{\gamma+1}} \ll 1, \quad  \lambda = 
\frac{2\pi}{(\gamma+1)} \frac{\eta p_1}{\theta M^{\frac{2}{\gamma+1}}}
 	\label{mu:lam}
 \end{equation}
This equation obviously possesses only one root, that belongs to the 
interval \( (\mu,1) \).  Since \( \lambda \gg \mu \) (\( \eta p_1 \) is 
always large in the case of the efficient solution) we have
\begin{equation}
	w \simeq \mu+\lambda \simeq \lambda
	\label{w1}
\end{equation}
in the case \( \lambda \ll 1 \).  In the opposite case \( \lambda \gg 1 \) 
we obtain
\begin{equation}
	w\simeq\frac{\lambda}{\lambda+1/(\gamma+1)}
	\label{w2}
\end{equation}
Both expressions    match at \( \lambda \sim 1 
\). Thus, the parameter \( \lambda  \) regulates the degree of shock 
modification for the efficient solution. In particular, for \( \lambda > 
1 \) one may write
\begin{equation}
	\frac{r}{r_{\rm s}}\simeq M^{\frac{2}{\gamma+1}}
	\label{r/rs}
\end{equation}
The subshock strength can in principle be reduced strongly in 
this case
\begin{equation}
	r_{\rm s} = \frac{4}{1+3 w^{8/3}} \simeq 1+\frac{3}{4}\frac{1}{\lambda}
		\label{r_s1}
\end{equation}
We have put \( \gamma=5/3 \) in to the R-H relation (\ref{c:r}), for short.  
However, \( \lambda \) itself depends on \( r_{\rm s} \) through \( \eta \) 
but remains bounded when \( r_{\rm s} \to 1 \).  Unfortunately, the 
existing injection theory essentially implies a finite subshock, so that it 
is difficult to judge the behavior of \( \eta \) at small \( r_{\rm s} - 1 
\).  As we shall see, \( r_{\rm s} -1 \) cannot, in fact, be very small, so 
that the dependence \( \eta(r_{\rm s}) \) is not so critical here.  We may 
rewrite eq.(\ref{r_s1}) as
\begin{equation}
	r_{\rm s}-1 \simeq 
	\frac{\theta}{\pi }\frac{M^{3/4}}{\eta (r_{\rm s}) p_1}
	\label{rs:min}
\end{equation}
Again one sees that formally the subshock can be quite weak for very large 
\( p_1 \).  Clearly, the details of acceleration in the regime \( r_{\rm 
s}-1 \la 1 \) may be quite sensitive to the injection mechanism at the weak 
subshock.  Generally, eq.(\ref{crit}) places both upper and lower limits on 
the parameter \( p_1 \).  To see how it works consider the limit \( \lambda 
\ll 1 \).  In this case the total compression ratio is almost independent 
of \( M \), whereas \( r_{\rm s} \) is close to four:
\begin{equation}
	\frac{r}{r_{\rm s}} \simeq \frac{3\pi}{4\theta}\eta p_1 \gg1, 
	\quad r_{\rm s} \simeq \frac{4}{1+3\lambda^{8/3}}
	\label{}
\end{equation}
(It is implied that at least \( \eta \sqrt{p_0 p_1} > 1\), see 
eq.(\ref{crit}), and therefore, \( \eta p_1 \gg 1 \)).  The subshock 
strength in this case is practically not affected despite a very strong 
modification of the flow in the precursor.  If we now raise \( p_1 \) so 
that \( \lambda \) becomes larger and \( r_{\rm s} -1 \) smaller, the 
condition (\ref{crit}) may be violated because of the large r.h.s.  
Denoting \( \eta(r_{\rm s} = 4) = \eta_0 \), and \( \eta_1 \) as the 
value of \( \eta \) at the smallest \( r_{\rm s} \) to satisfy the 
condition (\ref{crit}), we may express this condition in terms of \( p_1 \) 
as follows
\begin{equation}
	\frac{4B^2}{27p_0 \eta _0^2 } < p_1 < 
	\frac{1}{B}\left(\frac{\theta}{\pi}\right)^{\frac{3}{2}} 
M^{\frac{9}{8}}\sqrt{\frac{p_0}{\eta _1}}
	\label{crit2}
\end{equation}
This puts a limit on the Mach number below which the 
efficient-intermediate pair of solutions does not exist
\begin{equation}
M_{\rm cr}= \left(\frac{B}{3}\right)^{\frac{8}{3}}
\left(\frac{\pi}{\theta}\right)^{\frac{4}{3}}
 \left(\frac{4}{p_0 \eta_0^2}\right)^{\frac{8}{9}} 
\left(\frac{\eta_1}{p_0}\right)^{\frac{4}{9}}
	\label{M:cr}
\end{equation}
It can be seen that the magnitude of injection in both the cases of 
modified and unmodified subshock is critical for the existence of efficient 
solutions.  At the same time, the minimum strength of the subshock that is 
reached when \( p_1 \) approaches its upper limit in eq.(\ref{crit2}), is 
not very sensitive to the parameters:
\begin{equation}
	r_{\rm s\, min}-1 =B \sqrt{\frac{\pi}{\theta \eta_1 p_0}} 
	M^{-\frac{3}{8}}
	\label{rs:min2}
\end{equation}
The parameter \( \eta_1 p_0 \) is presumably rather small.  It 
characterizes the injection efficiency for a significantly modified 
subshock and preserves the subshock through quenching injection eventually.  
One sees that in practice the minimum subshock strength \( r_{\rm s\, min 
}-1 \) is of the order of unity since the Mach number is constrained by the 
condition \( \lambda \ga 1 \), or equivalently, \( M^{3/4} \la \eta p_1 \).  
\subsection{Flow profile in the CR precursor} The shock structure usually 
emerges from the balance between the nonlinear advection and the 
dissipation of the flow energy.  The dissipation is normally provided by 
viscosity or heat conduction or by both.  Heat conduction alone may or may 
not ensure a smooth shock transition whereas viscosity alway does, and a 
viscous subshock should be inserted whenever other transport phenomena like 
heat conduction cannot provide a smooth shock transition.  The shock 
amplitude and its thickness in media with quadratic nonlinearity are 
related by $ u l \sim \nu $, where $\nu $ is a relevant dissipation 
coefficient.

We started our consideration from these well known facts of shock wave 
theory, because a very similar situation arises in shocks strongly modified 
by CRs.  For example the relation $u_1 l \sim \kappa_1 $ holds in this 
case.  In contrast to the usual gasdynamics the flow energy is absorbed by 
the high energy particles, that have diffusivity $\sim \kappa_1 \equiv 
\kappa(p_1)$.  At least in the case of sufficiently high Mach numbers 
considered throughout this paper, inviscid smooth transitions are not 
possible in CR modified shocks for any finite \( p_1 \).  There are two 
reasons for that.  First, while $r_{\rm s} \to 1 $ the nonlinear shock 
structure associated with the acceleration process disappears as well ($r 
\to r_{\rm s}$, eq.(\ref{r1})).  Second, this structure also disappears if 
injection becomes less and less efficient as the subshock vanishes.

First, we consider the flow structure in a region where $x$ is not very 
small \( u_0^2 x/\Delta u\kappa_0 \gg 1 \), i.e.  basically beyond the 
diffusion length of injected particles.  Instead of eq.(\ref{fl:pot}) one 
may write
\begin{equation}
	\frac{d \Psi}{dx}=u_0+\sigma \sqrt{\frac{\kappa_0}{t_0}}
	\Phi\left(\sqrt{\Psi/\kappa_1^*}\right)
	\label{u:erf}
\end{equation}
where
\begin{equation}
	\Phi(x)=\frac{2}{\sqrt{\pi}}\int_{0}^{x}e^{-t^2}dt
	\label{}
\end{equation}
and \( \kappa_1^*=\kappa(p_1)/\theta \).
As in the previous subsection we focus on the efficient solution for 
which we obtain (see sec.5.1)
\begin{equation}
	\frac{d\Psi}{dx}=u_0+(u_1-u_0)\Phi\left(\sqrt{\Psi/\kappa_1^*}\right)
	\label{u:erf1}
\end{equation}
For \( \psi < \kappa_1^* \) we then obtain the following implicit solution 
for the flow potential \( \Psi(x) \)
 \begin{equation} 	\frac{\alpha^2}{2u_0}x=\frac{\alpha}{u_0}
	\sqrt{\Psi}-\ln\left(1+\frac{\alpha}{u_0}\sqrt{\Psi}\right)
 	\label{x(Psi)}
 \end{equation}
where
 \begin{equation}
 \alpha =2\frac{u_1 -u_0}{\sqrt{\pi \kappa_1^*}}
 	\label{alpha}
 \end{equation}
For \( \alpha\sqrt{\Psi}/u_0 < 1\) one obtains \( \Psi \simeq u_0 x+ 
(2\alpha/3 u_0)(u_0 x)^{3/2} \)
although this is correct only for \( u_0^2 x/\Delta u\kappa_0 \gg 1 
\).  The velocity profile in this region has the following form 
\begin{equation}
	u(x)=u_0+2\eta \Delta u \sqrt{\frac{p_0p_1}{\theta}}
	\sqrt{\pi\frac{u_0 x}{\kappa_0}}
	\label{u:sing}
\end{equation}
This singular behavior may only take place in the interval
\begin{equation}
	\frac{\Delta u}{u_0} < \frac{u_0 x}{\kappa _0} < 
	\left(\frac{u_0}{\Delta u}\right)^2\frac{\theta}{4 \pi \eta^2 p_0 p_1}
	\label{int:sing}
\end{equation}
that not necessarily exists. The prerequisite for this region 
to exist is a significant subshock reduction that is hardly possible as the 
preceding subsection suggests.
For smaller \( x  \) one has to use eq.(\ref{fl:pot}) that yields
\begin{equation}
	u=u_0+\pi \eta \sqrt{u_0 \Delta u}\sqrt{\frac{p_0 p_1}{\theta}}
	\frac{u_0 x}{\kappa_0}
	\label{u:x:sm}
\end{equation}
This was a thin region close to the subshock, related to the diffusion 
length of low-energy injection particles, whose dynamical role is assumed 
to be unimportant and no major change in flow speed occurs at this scale.  
It is perhaps worth mentioning that, if it were not so, the flow structure 
in this region would be strongly influenced by the details of injection 
which are not considered in this paper.

For larger \( x \), when \( (\alpha/u_0)\sqrt{\Psi} > 1\), the velocity
profile is again close to linear
\begin{equation}
	 u(x) \simeq 
	\alpha^2 x/2 +u_0\left(1+\ln \frac{\alpha^2 x}{2u_0}\right)
	\label{u:lin}
\end{equation}
which in the case of large \( u_1 \gg u_0 \) and for \( x < l \) can be 
simplified to
\begin{equation}
	u(x) \simeq u_1 \frac{x}{l}
	\label{u:lin:sim}
\end{equation}
where the total scale height of the precursor
\begin{equation}
	l=\frac{\pi \kappa_1}{2\theta u_1}
	\label{sc:h}
\end{equation}
This velocity behavior is valid for $u_0/u_1 < x/l <1$.  It is the most 
significant part of the shock transition.  It comprises not only the 
largest velocity variation.  It is also the most certain part regarding the 
asymptotic methods used here (see also below).  As mentioned earlier, 
besides this longest spatial scale there also exist two short scales 
associated with low energy injection particles (cf.  eq.(\ref{int:sing}))
\begin{equation}
 	l_1 = \kappa_0 \frac{\Delta u}{u_0^2}; \quad l_2=\kappa_0 
\frac{u_0}{\Delta u^2} \frac{1}{4 \pi \eta^2 p_0 p_1}
 	\label{l1:l2}
 \end{equation}
Far from criticality, i.e.  when \( \eta \sqrt{p_0 p_1} \gg 1 \) or when 
the subshock is not modified very strongly, the interval \( (l_1,l_2) \) 
collapses and these particles play no dynamical role in the shock 
structure.  Generally, this quasi-singular behavior does not seem to be 
persistent in the efficient solution.  However, this conclusion might 
change in the case of the inefficient or the intermediate solution and may 
be useful in interpretation of observations and simulations.  It should be 
recognizable as a region of very large \( du/dx \) just upstream from the 
subshock and generally should indicate that the acceleration process is 
close to criticality.  However, we shall not consider this situation in 
more detail here.

Beyond the precursor length, \( x > l \) from eq.(\ref{u:erf1}) we obtain
\begin{equation}
	u \simeq u_1 -\frac{u_1 -u_0}{\pi}\sqrt{\frac{2l}{x}} \exp 
\left(-\frac{\pi x}{2 l}\right)
	\label{u:as}
\end{equation}
Again, this region is less certain than that where $u_0/u_1 < x/l <1$,
since for $x \ga l$ the underlying asymptotic solution of the
diffusion-convection equation needs some minor modification (see
Sec.3). At the same time $u \simeq u_1 $ at this distance  and the
shock transition is virtually completed there. Furthermore, this
outermost part of the precursor is accessible only for particles that
in reality may begin to leave the system and whose behavior is
idealized to the greatest extent by the introduction of the sharp
cut-off at $p=p_1$. We see that there is no major physical reason for
improving formula (\ref{u:as}) in the present level of description.
\section{Particle spectrum}
The particle spectrum is the main goal in any acceleration scheme.  A 
formal expression for the particle spectral slope \(q=3 \beta \) was given 
already in Sec.3 (see eq.(\ref{g0})) in terms of the spectral function \( 
\bar V(p) \equiv V(p) +\Delta u\).  Its definition however involves \( 
\beta \) itself and, what is even more important, the flow structure across 
the whole shock transition.  Then an independent equation for \( V(p) \) 
was derived and solved, but the solution still contained the undetermined 
nonlinear eigenvalue \( A^* \) that we were able to specify only after the 
shock compression ratios had been obtained.  Moreover, the solution 
(\ref{c:d:sol}) to the diffusion-convection equation is strictly valid only 
under specific assumptions concerning the spectrum itself.  These 
assumptions can thus be justified only after the spectrum is calculated.  
The close link between all these quantities is a natural aspect of 
nonlinear shock acceleration.  It is this complexity of the acceleration 
process that has been standing in the way of a full analytic calculation of 
the particle spectrum for a long time.

Using the same normalization for \( g_0(p) \) as in eq.(\ref{Fb}), that
is \( g_0(p_0)=1 \), we obtain from eqs.(\ref{g0}) and  (\ref{J:d})
\begin{equation}
	g_0(p)=\frac{1}{J(p)} \exp\left[-\frac{3u_2}{\Delta u^*} \int_{p_0}^{p} 
	\frac{dp'}{p' J(p')}\right]
	\label{g0:1}
\end{equation}
For \( p \sim p_0 \) we have \(J\simeq1 \), and the particle distribution 
is \( g_0 \simeq (p/p_0)^{-3u_2/\Delta u} \), matching smoothly the thermal 
plasma at lower momenta via the solution of the injection problem, as it is 
explained in Appendix A. The same formula (\ref{g0:1}) is valid also for \( 
p_0 \le p \le p_1 \), while \( J(p) \) was given in Sec.4.  However, as 
explained, the solution for the function \( J(p) \) is strictly valid only 
when \( J(p) \gg 1 \), i.e.  for sufficiently large \( p \).  We seem to 
even have some difficulty in determining the magnitude of the spectrum in 
the momentum region where \(J(p) \gg 1 \), since \( J \) must be integrated 
in the exponent over the region where it is not determined accurately, i.e.  
in the region where \( J(p)-1 \sim 1 \).  \( J \) can be regularly 
calculated also in this case by continuing the \( \varepsilon- \) expansion 
in Sec.4.  At the same time this region is dynamically unimportant in the 
case of the efficient solution and there must be a way to get the high 
energy asymptotic result, i.e where \( J \gg 1 \) in eq.(\ref{g0:1}), 
without resolving this intermediate momentum range.  In fact we already 
exploited this approximation solving the integral equation (\ref{J}) for \( 
J \).  Namely, all necessary information about the behavior of \( J \) in 
the intermediate momentum range is `hidden' in the eigenvalue \( A^* \).  
cf.  eqs.(\ref{A*2}) and (\ref{omega}) we may rewrite eq.(\ref{g0:1}) as
\begin{equation}
	g_0(p)=\frac{1}{J(p)} \exp\left[-\frac{\Omega}{\sqrt{A^*}}+
\frac{3u_2}{\Delta u^*} \int_{p}^{p_1} 
\frac{dp'}{p' J(p')}\right]
	\label{g0:2}
\end{equation}
and for \( p \gg p_0 \) we may now use the solution (\ref{J(s)}) for \( J 
\).  Far from criticality (see Sec.5), when the efficient and the 
intermediate solutions are well separated, and focusing on the efficient 
solution, we have
\begin{equation}
	g_0(p) \simeq \frac{1}{J(p)}
	\label{g0:3}
\end{equation}
It is seen that for the efficient solution the softening of the spectrum in 
the region \( p \sim p_0 \) due to shock modification has no significant 
consequences for the high energy (\( p \gg p_0 \)) spectrum; $J$ rises so 
sharply that the exponential factor in eq.(\ref{g0:1}) remains $\simeq 1$.  
Using eqs.(\ref{s2}), (\ref{J:d}), (\ref{defs1}), (\ref{defs}), and 
(\ref{A1}), we then obtain from eq.(\ref{J(s)})
\begin{equation}
	J(p)=\frac{ \pi \eta \sqrt{\frac{p_1 
}{\theta}p}}{\sqrt{1+\frac{1}{3}\frac{\partial \ln J}{\partial \ln p} 
+\theta \frac{p}{p_1}}}
	\label{J:1}
\end{equation}
We observe that for \( p \ll p_1 \), \( J \simeq C\sqrt{p} \) satisfies 
this last equation; \( C \) is some constant.  One may also solve this 
equation exactly, transforming it to a linear equation for \( J^2 \).  
Indeed, introducing
 \begin{equation}
 	y=\frac{\theta^2}{\pi^2\eta^2p_1^2} J^2,\quad {\rm and} 
	\quad  \zeta =\theta \frac{p}{p_1}
 	\label{y:z}
 \end{equation}
we obtain
 \begin{equation}
 	\frac{1}{6}\frac{dy}{d \zeta}+\left(1+\frac{1}{\zeta}\right) y=1
 	\label{y:eq}
 \end{equation}
After simple calculations the solution that is regular at the origin can be 
written down in the following form
 \begin{equation}
 	y=\sum_{n=0}^{6} \frac{6! (-1)^n}{\zeta^n (6-n)! 6^n} 
-\frac{6!}{6^6 \zeta^6}e^{-6\zeta}
 	\label{y:sol}
 \end{equation}
This allows us to determine the constant \( \theta \). Since \( V_1 
\gg u_0 \) from eq.(\ref{defs}), we have
   \begin{equation}
   	\theta \simeq 1+\left.\frac{1}{3}\frac{\partial \ln J}{\partial \ln p}
	\right|_{p=p_1}
   	\label{tet1}
   \end{equation}
Which can be calculated to yield
   \begin{equation}
   	\theta =1+\frac{1}{6}\left. \frac{\zeta}{y} \frac{dy}{d \zeta}
	\right|_{\zeta=\theta} 
\approx 1.09
   	\label{tet2}
   \end{equation}
For practical purposes, instead of eq.(\ref{y:sol}), it is more convenient 
to use an approximate formula that follows immediately from eq.(\ref{J:1})
    \begin{equation}
    	J\simeq \pi \eta\sqrt{\frac{p_1}{\theta} p}\left(1+\beta +
\theta\frac{p}{p_1}\right)^{-\frac{1}{2}}
    	\label{J:2}
    \end{equation}
where \( \beta \simeq 1/6\) for \( p \ll p_1 \) and \( \beta =\theta - 1 
\), for \( p \sim p_1 \).  According to eq.(\ref{g0:3}) the particle 
spectrum at the shock front in this case is simply \( g_0 \simeq 1/J\) and 
the full spectrum can be given by
\begin{equation}
	g(x,p)=\frac{\sqrt{\theta}}{ \pi \eta \sqrt{p_1 p}}\left(1+\beta
+\theta\frac{p}{p_1}\right)^{\frac{1}{2}} \exp \left[-\frac{1+
\beta}{\kappa(p)}\Psi (x)\right]
	\label{g(x,p)}
\end{equation}
Clearly the spectrum hardens with the distance upstream and it is well 
localized at the upper momentum cut-off for \( \Psi/\kappa(p_1) \ga 1 \). 
The most interesting aspect of this solution is that the postshock 
spectrum \( g_0(p) \) is a power-law in a broad momentum range with a 
universal index coinciding with the test particle result for the shock 
of compression ratio~7. For \( p   \) approaching \( p_1 \), the spectrum 
\( g_0 \) flattens. Nevertheless, its slope remains universal, decreasing 
from 1/2 for \( p_0 \ll p \ll p_1 \) to about 0.3 at \( p=p_1 \), 
independently of any parameters.
\section{Discussion and conclusions}
In this paper we have found that kinetic solution for diffusive shock 
acceleration which emphasizes a very strong coupling of accelerated 
particles with the gas flow.  We have also demonstrated that there are 
three different solutions in a specified region of parameter space.  This 
region is undoubtedly of great astrophysical interest.  Therefore, the 
question, how efficient shock acceleration is, implies automatically the 
question which solution among these three is the attractor of a system 
evolving in time.  This question can be also put in the following context.  
Is shock acceleration a passive process in which only a limited fraction of 
the shock energy is transformed into more or less parasitically accelerated 
particles, or is the shock itself radically restructured by these particles 
and they consume almost all its energy? Are transitions between these 
different states possible and if so, how do they occur? Thus, the notion of 
critical phenomena appears to be quite plausible for nonlinear shock 
acceleration.

The first step of any quantitative theory of such phenomena should be the 
identification of a system parameter and the determination of its critical 
values.  This parameter (or parameters) should encompass all the physically 
relevant quantities.  As it was shown in the present paper, the most 
important system parameter is $\Lambda =\eta p_1 M^{-3/4}$.  For instance, 
if we consider the case in which \( M \) is fixed and sufficiently large, 
so that $ \Lambda < 1$, acceleration is essentially governed by the 
parameter \( \Lambda_1 =\eta p_1 \) (see sec.5).  In particular, the total 
compression \(r \propto \eta p_1 \) in this case.  Other regions in \( 
\Lambda_1, M \) parameter space are discussed in sec.5, and one important 
statement is very simple: there always exists a critical \( 
\Lambda_1^{(1)}(M) \) such that for \( \Lambda_1 < \Lambda_1^{(1)} \) only 
the inefficient solution is possible.  Moreover, there is an upper critical 
value for \( p_1 \), or more precisely for \( p_1 \eta_1^{1/2}\) (see 
eq.(\ref{crit2})), beyond which the efficient solution fails to exist 
again.  The reason is the following.  Once this parameter exceeds the 
critical value, the backreaction of the CRs on the foreshock flow becomes 
so strong and, as a result, the spectrum at lower energies so steep, that 
the required CR pressure cannot be maintained.  Also at \( \Lambda_1 > 
\Lambda_1^{(1)} \), the system was found to be in quite different states, 
ranging from \( r \propto \Lambda_1, \, r_{\rm s} \simeq 4 \) independent 
of \( M \) (when \( \Lambda \ll 1 \)), to \( r \propto M^{3/4} \) with a 
noticeable subshock reduction \( r_{\rm s} -1 \sim 1 \), in the opposite 
extreme \( \Lambda > 1 \).

This regime (\( r \propto M^{3/4} \)) could basically be quite natural also 
for the \tfm\ {\nolinebreak(\cite{als};} \cite{dru:vlk}) since it corresponds 
effectively to the case \( \Lambda_1 = \infty \) (in the \tfm\ it is 
implied that $p_1 \to \infty $), and once the subshock is modified, one 
could very easily estimate also the total shock modification as 
(eq.(\ref{c:r})) \( r \propto M^{3/4} \).  It is important to emphasize 
that such an estimate results from only two assumptions.  The first says 
that the subshock is noticeably modified, and the second that it is not 
smeared out completely.  No further calculations are needed.  It means 
unambiguously that under these circumstances the compression ratio of a 
steady large Mach number shock must be rather large.  In the \tfm, however, 
\( r_{\rm s} \) always crosses unity in the case of efficient solution 
already at \( M \sim 10 \) even if \( \eta \to 0 \).  \footnote{It is 
noteworthy that the sequence of limiting transitions is important here, 
i.e.  in the process $ p_1 \to \infty $, $\eta \to 0$ the parameter 
$\Lambda_1 $ remains infinite.  We recall in this regard that the problem 
under question is a singularly perturbed one in the parameter $1/p_1$.} One 
is forced to take that branch of two-fluid solution of R-H relations in 
which \( r_{\rm s} \equiv 1\) beyond this critical \( M \) and the scaling 
\( r \sim M^{3/4} \) cannot be recovered since it would correspond to a 
rarefaction wave \( r_{\rm s}<1 \).  In contrast, the inequality \( r_{\rm 
s} > 1 \) is always guaranteed in the kinetic description for any finite 
$\Lambda_1$.  This is a fundamental difference between kinetic and 
hydrodynamic descriptions.  The latter is not capable of preserving the 
kinetic link between injection and high-energy particles.  It allows \( 
r_{\rm s} = 1 \) which means on the kinetic level of description an 
infinitely steep suprathermal particle distribution for \( p \ga p_0 
\)\footnote{This argument is not correct when the gas flow develops a very 
large gradient just in front of the subshock due to the pressure of 
suprathermal particles.  Similar situation was briefly considered in 
Sec.5.2.}.  This must normally result in \( P_{\rm c} \to 0 \).  However, 
this does not occur within the two-fluid description in which CRs and 
thermal plasma are coupled only hydrodynamically, not kinetically.  In the 
kinetic picture presented above, this link is established mathematically 
through the nonlinear integral equation that simply has no eigenvalues 
corresponding to efficient solutions when the subshock is too weak and the 
spectrum in the suprathermal region too steep with the implication that the 
CR-pressure does not suffice.  The kinetic link of suprathermal and high 
energy particles is a constitutive aspect of selfregulation in nonlinear 
shock acceleration.  Indeed, it leads to the following universal (Mach 
number independent) relation between the flow deceleration in the smooth 
and in the discontinuous parts of the shock transition (eq.(\ref{r1}))
\begin{equation}
	\frac{u_1-u_0}{u_0-u_2}=\frac{\pi}{\theta} \eta p_1
	\label{rh:m}
\end{equation}
Therefore, the modification effect disappears whenever does the subshock.  
In fact this happens even earlier, since the criterion (\ref{crit}) will be 
violated and this solution disappears abruptly when \( u_0-u_2 \) drops 
below some critical value (eq.(\ref{rs:min2})).

One remarkable result that came about concerns the downstream spectrum \( 
g_0(p) \).  Over a wide portion of momentum space the power-law index 
depends on absolutely nothing! It equals \( 3\frac{1}{2 } \) for the 
standard normalization of particle distribution.  Interestingly, this 
coincides precisely with a test particle spectrum in an unmodified shock of 
compression ratio~7.  Thus, if the power-law index would be calculated 
using the ratio of specific heats \(\gamma_{\rm c} \) (which is clearly 
very close to 4/3) rather than the compression ratio, one would obtain 
essentially the same result.  Our solutions imply that \( r \) can be 
large, at least larger than 7.  Nevertheless the spectrum fails to flatten 
with growing compression.  The resolution of this `paradox' lies obviously 
in the thickness of the shock, that grows simultaneously with \( r \) and 
\( p_1 \), and 
these two oppositely acting factors compensate each other exactly.  
Similarly there is no flattening of the spectrum with momentum unless \( p 
\sim p_1 \), since particles that sample larger compression at larger 
momenta see a more extended flow structure as well.  One could say that 
while in the test particle theory the energy spectrum is scale invariant 
because all the internal shock scales are irrelevant, in the nonlinear 
theory both the momentum spectrum and the shock structure are scale 
invariant in a certain region of phase space.  The spectrum flattening at 
the cut-off (\( q \) decreases from 3.5 to about 3.3) is obviously caused 
by the presence of this sharp cut-off itself.  The mathematical nature of 
this flattening is very simple and can be explained using the analytic 
properties of the spectral function \( J(p) \) and hence, those of \( 
g_0(p) \) in the complex \( p \)-plane.  Namely, in an unphysical region at 
\( p=-p_1 \), \( g_0 \) has the branch point, which is formally responsible 
for the flattening at \( p=p_1 \).  In a physically more realistic case in 
which the losses are distributed at \(p \sim p_1 \), this flattening may 
not appear.  (We therefore do not speculate for example about a possible 
link of this spectrum flattening with the bump around the energy $ 10^{14} 
eV$ on the galactic CR spectrum.) In the present picture, however, 
particles with \( p \la p_1 \) seem to `know' how high is the cut-off, 
which appears to be strange from the point of view of causality.  This 
`information' is provided again, by the flow structure, in fact through the 
singularity of the spectral function \( V(p) \) at \( p=-p_1 \).  As 
particles come nearer to \( p_1 \) they also diffuse to the outer region of 
the precursor where \( u(x) \) approaches \( u_1 \) exponentially.  Hence 
these particles develop a spectrum that is closer to $q=3r/(r-1)$.  Recall 
that in the case under consideration \( r \gg 1\).  Moreover, if we 
formally extend \( g_0(p) \) beyond \( p=p_1 \), considering $g_0$ as a 
test particle spectrum, we get the power-law index \( q=3 \) as \( p \to 
\infty \) instead of the `exact' \( q=3/(1-r^{-1}) \simeq 3 \).  This 
small deviation can be easily removed and the modification needed for that 
will not affect the region where \( q=3\onehalf \).  This difference in the 
spectrum slope occurred because, for simplicity, we treated the dynamically 
unimportant region \( \Psi > \kappa_1 \) on the same basis as the region \( 
\Psi < \kappa_1 \) (see the end of sec.5).

It would be interesting to examine what may happen in a time dependent 
case.  Formally, the generalization of our results to time dependent 
acceleration should not be a serious problem.  Namely, for sufficiently 
large \( p_1 \) one can adopt an `adiabatic' approximation in which the 
solution has approximately the same form as that found in this paper, 
except for parameters this solution depends on; they will slowly evolve 
with time.  If the Mach number can be fixed, the only parameter we need is 
\( p_1(t) \), all the others and the flow configuration are calculable.  
The cut-off increment may, at least tentatively, be obtained from the usual 
equation \( dp_1/dt =p_1/\tau_{\rm acc}(p_1) \).  Since the acceleration 
time equals \( \tau_{\rm acc} \sim \kappa(p_1)/u^2 \), the process should 
be slow compared with the time scale at lower momenta.  This should justify 
an adiabatic approach.  However, the spectrum at the cut-off may differ 
from our steady state solution significantly.

Our main conclusion concerning the overall spectrum to have a universal 
index $q=3\onehalf$ may also be a subject for corrections due to a number 
of physical reasons ignored in our consideration.  These should be taken 
into account before we attempt to compare our results with any observations 
or numerical calculations.  We have studied the simplest case of a plane 
steady shock with specified Bohm diffusion in which the magnetic field 
either plays no dynamical role or is quasiparallel to the shock normal.  
Clearly a more realistic geometry, time dependence and losses increasing 
continuously with the particle energy, will steepen the spectrum.  We also 
have simplified our description of subrelativistic particles replacing \( 
p(1+p^2)^{-1/2} \) by unity in the pressure integral (eq.\ref{Fb}), (see 
eq.(\ref{J}) and the text below it).  Doing so we could overlook some 
feature in the particle spectrum at \( p \sim 1 \) and ignore the role of a 
possible relativistic peak at \( p \simeq 1 \) in the partial pressure.  
Generally, this peak may or may not develop in a modified shock.  In the 
cases considered throughout this paper its contribution to the particle 
pressure would be much smaller than that of the high energy particles \( p 
\la p_1 \) (see Appendix A in \cite{mv96}).  The most probable correction 
to the particle spectrum due to the relativistic peak should be a turn in 
the spectral slope at \( p \sim 1\) just because of the correspondent 
factors in the kernel of the integral equation (\ref{i:eq:p}).  And 
finally, particle diffusion is normally assumed to be provided by waves 
that, in turn, are excited by accelerated particles.  This link may not 
only soften the spectrum but also change other characteristics of 
acceleration process noticeably.  Notwithstanding the physical importance 
of the above factors, we believe that the key issue for understanding the 
nonlinear shock acceleration is the gas flow--CR coupling.  This coupling 
may, at least formally, operate under prescribed injection, diffusivity and 
upper cut-off momentum.  It should be understood as such before a 
comprehensive nonlinear theory of shock acceleration is done.  The major 
reason for that is of course the multiplicity of the solution.

We have presented only the steady state solutions and specified the 
conditions under which they appear multiply. The main issue now is of 
course their realisability. Whereas for the case of a unique solution a 
proof of 
its stability suffices, in a parameter range where 
all the three solutions are admitted, the full time dependent 
calculations may be needed to predict the system behavior.

Already at this stage several important features of the acceleration 
process can be pointed out.  First, there are no such `minor' things like 
the injection rate or the cut-off momentum that can be considered as having 
a small effect on the acceleration process and their treatment should not 
be oversimplified just by arguing that the former is small and the latter 
is large.  These parameters enter all the main results and in many cases 
symmetrically, \eg through \( \Lambda_1 = \eta p_1 \).  Therefore, both \( 
\eta \) and \( p_1 \) are equally important.  This conclusion is strictly 
valid for a steady state.  One can imagine a time evolution in which \eg \( 
p_1 \) increases but \( \eta \) decreases leaving a parameter such as \( 
\Lambda_1 \) that governs the steady state flow structure (or its 
equivalent in a time dependent formulation, which would have the meaning of 
the energy content of accelerated particles) approximately constant.  
Furthermore, in time dependent situations the constraint (\ref{rh:m}) is 
strictly speaking not valid, and a complete smoothing of the subshock 
cannot be excluded.  This last remark might be important for the 
interpretation of numerical results.

A number of time dependent numerical solutions were obtained in the past 
(\eg \cite{bell87}; \cite{fg87}).  Except perhaps the case considered by 
Bell, (\( M=100) \) it is not clear whether they are within the parameter 
range where multiple solutions appear or should rather be regarded as 
inefficient solutions, sometimes with quite a strong shock modification and 
high acceleration efficiency.  Furthermore, at moderate Mach numbers and 
near the bifurcation point the `inefficient' solution may be not very 
different from two solutions that we termed efficient and intermediate.  We 
will discuss the relations between these three solutions in more detail in 
Paper II.

\acknowledgments{I wish to express special thanks to Heinz
V\"olk for drawing my attention to the problem of nonlinear shock
acceleration, for many fruitful discussions of this subject, for his
thoughtful reading of this manuscript and for his invaluable comments
and suggestions concerning this paper. This work was done within the
Sonderforschungsbereich 328, ``Entwicklung von Galaxien'' of the
Deutsche Forschungsgemeinschaft (DFG).}

\appendix
\section{Injection}
The injection rate calculated in (MV95, M96, MV97), essentially depends on 
details of the subshock dissipation mechanism like the level and the 
spectrum of underlying plasma turbulence.  The obliqueness of the magnetic 
field and the presence of reflected particles (\cite{sch90}) will clearly 
influence the injection rate too.  These parameters may vary from model to 
model and it is therefore perfectly reasonable to consider the injection 
rate \( \eta \) as another free parameter along with \( M \) and \( p_1 \), 
at least in the context of nonlinear shock acceleration studied here.  The 
subshock compression ratio \( r_{\rm s}(\eta,M,p_1) \) obtained in this 
paper may then be combined with the injection rate \( \eta =\eta (r_{\rm 
s}) \) obtained in the papers listed above to yield both \( r_{\rm 
s}=r_{\rm s}(M,p_1) \) and \( \eta = \eta (M,p_1) \).  However, a 
microscopic subshock dissipation mechanism that operates 
quasi-independently of the large scale shock modification, should be 
identified for this purpose in each particular physical model (see M96 for 
an example of such a mechanism).

Besides the injection rate \( \eta \) we introduced the injection momentum 
\( p_0 \) (eq.(\ref{P_c})) in order to separate the CR and the thermal 
populations.  This can be done unambiguously if the injection is weak so 
that the injected particles which are between these two populations in the 
momentum space, do not modify significantly the flow upstream by themselves 
but only after being accelerated to very high energies and therefore over a 
much larger scale.  This is the essence of our approach to distinguish 
between thermal and high-energy populations based on the difference in 
diffusion length of the respective components.  Clearly the monotonic 
growth of the diffusion coefficient with the particle momentum is very 
critical here.  If the subshock is sufficiently strong, these two 
populations of particles do not even overlap in the upstream phase space, 
Fig.2.  They are separated by the region \( v_0 < \left|{\bf p}/m \right| < 
\alpha (u_0-u_2) \), where \( \alpha \ga 1\) (M96), \( \bf p \) is measured 
in the local upstream frame where \( u(x) \simeq u_0 \) and \( v_0 \) is 
the thermal velocity at the same point.  Therefore, the momentum \( p_0 \) 
is a rather formal lower boundary separating the CR gas from thermal plasma 
and from the viewpoint of the upstream distribution, \( p_0 \) (or may be 
\( p_0(\theta) \), where \( \theta \) is the pitch-angle) can be chosen 
arbitrarily within this empty part of the phase space.  Turning to the 
downstream region we note that the injection solution yields the particle 
spectrum that starts from the thermal Maxwellian and evolves at higher 
momenta into the power-law determined by the subshock compression.  This 
allows the smooth matching of this injection solution with the low-momenta 
asymptotic solution of eq.(\ref{c:d}) within an extended overlapping 
interval in the region \( p \ga \alpha m(u_0-u_2) \), (see also M96, MV97).  
This means that again, one should be able to choose \( p_0 \) arbitrarily 
within the overlapping interval and all physically meaningful results must 
be independent of \( p_0 \).  \footnote{This statement should not be taken 
literally.  First of all, \( p_0 \) formally enters the definition of the 
CR number density \( n_{\rm c} \) (sec.4) and thus the parameter \( \eta 
\).  But this is an implicit dependence on \( p_0 \) through \( n_{\rm c } 
\) and once \( n_{\rm c } \) is defined \( p_0 \)-invariant, (\eg by 
subtraction of the thermal Maxwellian from the total particle 
distribution), this dependence disappears.  Second, the ``injection 
momentum'' \( p_0 \) can appear in some criticality conditions (Sec.5) and 
also in situations when the approximation described just below is only 
marginally correct.  This means that the description of the injection 
process by only one parameter \( \eta \) is insufficient in such a 
situations.  On the other hand in selfconsistent injection models both the 
parameters \( \eta \) and \( p_0 \) should be explicitly specified in terms 
of physical quantities like the thermal velocity and the Mach number.  If 
so, a complete solution of the acceleration problem depends not on 
parameters like \( \eta \) or \( p_0 \) but only on these physical 
quantities.  At the same time, as it may be seen from Fig.2, the transition 
region between the thermal Maxwellian and the high-energy power-law is 
relatively narrow, so that the quantity \( p_0 \) has also a clear physical 
meaning.} If \( p_0 \) indeed lies in this empty phase space region 
upstream, then \( P_{\rm inj} =0\) in eqs.(\ref{m:c:p},\ref{Fb}).  However, 
such a choice of \( p_0 \) may cause technical difficulties since the 
diffusion-convection equation is not valid in this region due to anisotropy 
of pitch-angle distribution.  If we take \( p_0 \) larger, just to make the 
diffusion-convection equation valid, then \( P_{\rm inj} \neq 0\) and must 
generally speaking be retained in Bernoulli's integral, 
eqs.(\ref{m:c:p},\ref{Fb}).  It is also worth noting that the matching of 
the downstream injection solution onto the standard power-law occurs 
typically at \( p \) somewhat larger than the void in the upstream 
distribution.  Thus, \( p_0 \) may be slightly higher than the lower 
boundary of the upstream CR distribution.
 
The above details of the phase space geometry, being essential for 
injection and for its link with the further acceleration, are not so 
critical for the subject of this paper, i.e.  for the shock modification by 
high-energy particles.  It should be emphasized, however, that an important 
physical condition for legitimating both the injection theory and the 
present study is that the suprathermal particles with momenta \( p \sim p_0 
\), play no considerable dynamical role in the precursor, i.e.  their 
contribution to \( P_{\rm c} \) is much smaller than that of the rest of 
momentum space \( p_0 \ll p \le p_1 \).  As we have stressed already, they 
do not produce a noticeable flow variation over their own diffusion length 
\( \kappa(p_0)/u_0 \) which for example justifies the adoption of the 
injection calculations performed in a quasi-homogeneous upstream flow.

The injection theory produces a suprathermal downstream asymptotics \( 
g_0(p) \simeq Qp^{-q_0} \) where \( q_0=3u_2/(u_0-u_2) \) and \( Q \) 
depends also on \( q_0 \) along with a number of other plasma parameters 
near the subshock like the downstream temperature and the magnitude of 
MHD-turbulence (see MV95, M96).  This is essentially the same as 
eq.(\ref{low:m}).  The theory of injection also determines the spatial 
distribution of \( g(x,{\bf p}) \) at the subshock for particles with 
momenta \( \kappa(p)/u_0 \sim l_{\rm inj} \equiv \kappa (p_0)/u_0 \ll l \), 
where \( u(l) \simeq u_1 \).  Therefore \( l_{\rm inj} \) is virtually 
irrelevant for our further consideration and only the parameter \( Q \) is 
needed.
\section{Diffusion-convection equation}
We examine the idea that the result (\ref{ren:g}) that is strictly valid 
for small values of \( \Psi/\kappa \), is also a good approximation to the 
solution of the convection diffusion equation in general.  To this purpose, 
it is natural to seek the solution of eq.(\ref{c:d}) in the following form
\begin{equation}
	g=g_0(p)\exp \left\{-\frac{1+\beta}{\kappa}\Psi\right\}\hat g(p,\Psi)
	\label{ren:g:p}
\end{equation}
Clearly we have \( \hat g(p,0)=1 \) as a boundary condition.  Our objective 
here is to demonstrate that \( \hat g(p,\Psi) \simeq 1 \) also for \( 
\Psi/\kappa \) noticeably larger than unity.  A practically important 
region is of course that where \( \Psi/\kappa \) is not very large due to 
the exponential factor in eq.(\ref{ren:g:p}).  Introducing also $$ Q(\Psi) 
\equiv\frac{d}{d\Psi} \ln u(\Psi), $$ for the function \( \hat g \) we have 
the following equation that can be easily derived from eq.(\ref{c:d})
\begin{equation}
	Q(\Psi)\left(\frac{\partial \hat g}{\partial \tau} 
	-\kappa\frac{\partial \hat g}{\partial \Psi} -\Psi\hat g 
	\frac{\partial}{\partial \tau} \frac{\beta +1 
	}{\kappa}\right)=-(1+2\beta)\frac{\partial \hat g}{\partial \Psi} 
	+\frac{\beta}{\kappa} (1+\beta) \hat g +\kappa \frac{\partial^2\hat 
	g}{\partial \Psi^2}
	\label{g:hat}
\end{equation}
where \( \tau \equiv 3 \ln p/p_0 \).  At this point we need some more 
information about the function \(Q(\Psi) \).  As we have shown in sec.5.2., 
for not very small \( \Psi \), more precisely for
\begin{equation}
	 \Psi/\kappa_0 > 
	(u_0/\Delta u)^2/\eta^2p_0 p_1,  
	\label{l:ineq}
\end{equation}
and in the case of strong shock modification,
the velocity of the flow behaves as follows (see eq.(\ref{u:erf}))
\begin{equation}
	u(\Psi) = u_1 \Phi\left(\sqrt{\frac{\Psi}{\kappa (p_1)}}\right)
	\label{dpsi:dx}
\end{equation}
where \( \Phi \) is the probability integral.  Thus, for \( \Psi \la 
\kappa(p_1) \), to a good approximation we may write
\begin{equation}
	Q(\Psi)\simeq \frac{1}{2\Psi}
	\label{Q}
\end{equation}
For larger \( \Psi > \kappa_1 \) the function \( Q(\Psi) \) falls off 
exponentially.  But this region is dynamically unimportant (\( g \) is 
exponentially small already for \( p \la p_1 \)) and we confine our 
consideration to the case in which eq.(\ref{Q}) holds.  Introducing a new 
variable \( z=\sqrt{2\Psi/\kappa} \) and performing the variable 
transformation \( (\Psi,\tau) \mapsto (z,\tau)\) eq.(\ref{g:hat}) rewrites
\begin{equation}
	\frac{\partial \hat g}{\partial \tau} +\sigma(\tau) z \frac{\partial 
	\hat g }{\partial z} + \delta (\tau) z^2\hat g -\frac{\partial^2 \hat 
	g}{\partial z^2}=0, \quad \hat g(\tau,0)=1
	\label{g:hat:1}
\end{equation}
where 
\begin{equation}
	\sigma=\frac{5}{6}+2\beta(\tau)  \quad {\rm and} \quad 
\delta=(1+\beta)(\frac{1}{6}-\beta)-\frac{1}{2}\frac{\partial 
\beta}{\partial \tau}
	\label{delta}
\end{equation}
Physically, the problem (\ref{g:hat:1}) should be posed as follows.  At \( 
\tau \simeq 0 \,(p \simeq p_0), \, \hat g(z,\tau) \) has to be matched 
smoothly onto the solution of the injection problem as it was discussed 
earlier.  Then \( \hat g \) and hence \( g \) can be determined from 
eq.(\ref{g:hat:1}) provided that \( \beta(\tau) \) is known.  This latter, 
in turn, depends on the spectral index of the downstream distribution to be 
found from eq.(\ref{b:c}).  As it is shown in Appendix C, unless \( p \sim 
p_0 \) or \( p \sim p_1 \) the function \( \beta(\tau) \) and thus \( 
\delta(\tau) \) is a slow function of \( \tau \).  We observe that a 
characteristic `time scale' in eq.(\ref{g:hat:1}) is \( \tau \sim 1 \).  
Thus, taking \( \sigma \) and \( \delta \) as approximately constant one 
can easily solve the problem (\ref{g:hat:1}) in terms of parabolic cylinder 
functions.  But even this simple solution is not really needed for our 
purposes.  It turns out (Appendix C) that \( \delta \ll 1 \) for \( p_0 \ll 
p \la p_1 \).  Consider the region \( \tau >1 \), where the solution 
`forgets' the `initial' condition \( \hat g(\tau\simeq 0,z ) \), that is 
given by the high energy end of the injection solution.  The solution to 
eq.(\ref{g:hat:1}) will be determined by the boundary condition \( \hat 
g(\tau,0)=1 \) and one may see that for not too large \( z <1/\sqrt{\delta} 
\) we also have \( \hat g(\tau,z) \simeq 1 \).  Furthermore, the injection 
solution yields the spatial distribution of the particle spectrum that is 
consistent with the exponential decay \( \propto \exp 
(-(1+\beta)\Psi/\kappa )\) (see MV95).  We may thus put \( \hat g(0,z)=1 \) 
which yields \( \hat g(\tau,z)\simeq 1 \) for small \( \delta \).  As we 
mentioned above this solution is not valid for \( z \ga 1/\sqrt{\delta} \), 
but once \( \delta \) is small we can ignore this region since \( g(p,z) 
\sim g_0 \exp(-1/2\delta ) \) there.

Finally,  we note that this simple solution (\( \hat g = 1 \)) of
eq.(\ref{g:hat:1}) is exact if we require \( \delta \equiv 0 \) in
eq.(\ref{delta}).  This yields for the spectral index
\begin{equation}
	q= 3\beta(\tau)=3\frac{1+\beta_0+(6\beta_0 -1)e^{-7\tau/3}}{6(1+\beta_0) 
-(6\beta_0 -1)e^{-7\tau/3}}
	\label{3bet}
\end{equation}
where \( 3\beta_0 \) is the spectral index at \( \tau \simeq 0, \, (p 
\simeq p_0 \)), which is given by the injection theory and coincides 
with the standard power-law index calculated for the subshock compression 
ratio.  For larger \( \tau \), \( \beta \) drops to 1/6 rapidly.  Clearly, 
for \( \hat g =1 \) to hold, the above behavior of \( \beta(\tau) \) must 
be consistent with the solution of eq.(\ref{b:c}), which will be confirmed 
in Appendix~C.
\section{Verifying assumptions}
It has been understood for a long time that in contrast to the test
particle theory, the shock acceleration in the nonlinear regime cannot
be decomposed as a sequence of  independent processes like the
formation of the overall flow structure, particle spectrum, and
injection, or losses. Even the magnitude of the cut-off momentum cannot
be specified prior to obtaining the full solution when it is influenced
by the solution itself (we did not consider this case here).  All these
factors are linked very tightly in the solution obtained in the
preceding sections and non of them can be prescribed a priori.
Nevertheless, we made several assumptions concerning mostly the flow
profile and the particle spectrum. For the sake of convenience we list
here these assumptions along with the confirming results.

First, we neglect the adiabatic gas compression term in the Bernoulli's 
equation upstream, eq.(\ref{m:c:p}). Denoting 
\begin{equation}
	v=\frac{u(x)}{u_1} \quad {\rm and} \quad  {\mathcal{F}}(x)=1-
	\frac{1}{\rho_1 u_1^2} \left(P_{\rm c}+P_{\rm inj}\right)
	\label{}
\end{equation}
we rewrite eq.(\ref{m:c:p}) as
\begin{equation}
	F(v) \equiv v+\frac{1}{\gamma M^2}\left(\frac{1}{v^\gamma}-1\right)
	={\mathcal{F}}(x)
	\label{ber:v}
\end{equation}
We must specify the condition under which we may invert \( F(v) \) to
obtain \( v \simeq {\mathcal{F}} \).  First of all we observe that the
l.h.s.  of eq.(\ref{ber:v}) has a very sharp minimum at \( v=v_{\rm
m}=M^{-2/(\gamma +1)} \).  Since \( v_{\rm m} \ll 1 \) the
approximation \( v \simeq {\mathcal{F}} \) may be good over most of
the precursor, where \( v > v_{\rm m} \) (see Fig.3).  Indeed, using
eq.(\ref{u:x:sm}) we can rewrite this inequality as
\begin{equation}
	v =\frac{u_0}{u_1} \left[1+\pi \eta \sqrt{\frac{p_0 
p_1}{\theta}} \sqrt{1-r_{\rm s}^{-1}}\frac{u_0 x}{\kappa_0} \right] > 
M^{-\frac{2}{\gamma +1}}
	\label{constr1}
\end{equation}
In the trivial case \( (u_1/u_0) M^{-2/(\gamma +1)}  
 \ll 1\), i.e. \( \nu u_1/u_0 \equiv 2M^{-2}(u_1/u_0)^{\gamma+1}\ll 1\) 
the approximation \( v \simeq 
{\mathcal{F}} \) is correct for all \( x > 0 \). If only the weaker 
condition \( \nu \ll 1 \) is fulfilled, we obtain the following constraint 
(eqs.(\ref{r1},\ref{constr1}))
\begin{equation}
	\frac{u_0 x}{\kappa_0} \gg \sqrt{(1-r_{\rm s}^{-1})\frac{p_1}{p_0}} 
	M^{-\frac{3}{4}}
	\label{constr2}
\end{equation}
Substituting \( p_{1 \, {\rm max} } \) and \( r_{\rm s\, min} \) from 
Sec.5.1, we obtain
\begin{equation}
	\frac{u_0 x}{\kappa_0} \gg M^{-\frac{3}{8}}(\eta_1 
)^{-\frac{1}{2}} \sim (r_{\rm s \, min} -1) \sqrt{p_0}
	\label{constr3}
\end{equation}
That means that we may neglect the adiabatic compression term already
inside the scale height of injected particles \( \sim \kappa_0/u_0 \).
Smaller \( x \) belong virtually to the internal subshock structure
which cannot be resolved within the diffusion-convection description in
any case. If the nonrelativistic correction to the diffusion
coefficient is important, one should write \( \kappa_0=\hat\kappa_0/p_0
\), where \( \hat\kappa_0 \) is the `true' diffusion coefficient at \(
p=p_0 \). Then the inequality (\ref{constr3}) becomes more stringent
and \( x \) must be outside the diffusion length of low energy injection
particles.

Our next assumption concerned the functions \( \beta (\tau) \) and \( 
\delta (\tau) \) that enter eq.(\ref{g:hat:1}).  According to Sec.6 and for 
sufficiently large \( p \gg p_0 \) we may write \( \beta \simeq (p/3)( 
\partial \ln J/ \partial p) \), to yield
\begin{equation}
	\beta \simeq \frac{1}{6} \cdot \frac{1}{1+6\theta p/7p_1}
	\label{}
\end{equation}
One sees that \( \beta \) is close to \( 1\over 6 \) for \( p \) not too 
close to the upper cut-off.  What is even more important for our treatment 
of eq.(\ref{g:hat:1}) in Appendix B is that $ \delta \ll 1$.  Even in the 
worst case possible, namely when $ p \to p_1 $, we still have $\delta \la 
0.1 $, so that the result (\ref{c:d:sol}), based largely on the above 
behavior of parameters $\beta $ and $\delta $, is still a good 
approximation to the solution of the diffusion-convection equation.

A number of our approximations depend on the value $V_0 \equiv V(p_0) $, 
Sec.3.1.  Unfortunately it is impossible to determine this quantity 
directly from our asymptotic solution for $V(p) $ since it is formally 
valid for $V(p) \gg V_0 $.  We therefore should estimate $V_0$ directly 
from the definition (\ref{V}).
\begin{equation}
	V_0 \sim \left . \frac{du}{dx} \right |_{x=0} \frac{ \hat \kappa_0}
{u_0}
	\label{}
\end{equation}
The quantity $ \hat \kappa_0 = p_0 \kappa_0 $ accounts of the 
nonrelativistic correction to the diffusion coefficient used.  According to 
eq.(\ref{constr1}) we thus have
\begin{equation}
	\frac{V_0}{u_0} \sim  \pi \eta \sqrt{p_0 p_1} p_0 \sqrt{1-1/r_{\rm s}}
	\label{v:ov:u}
\end{equation}
Since the criticality  parameter $ \eta \sqrt{p_0 p_1} > 1$ cannot be
very large and $\Delta u $ not very small (see the end of Sec.5.1),
we infer that $V_0 < \Delta u$ is a reasonable approximation for
sufficiently small $p_0$.

\newpage
	\figcaption[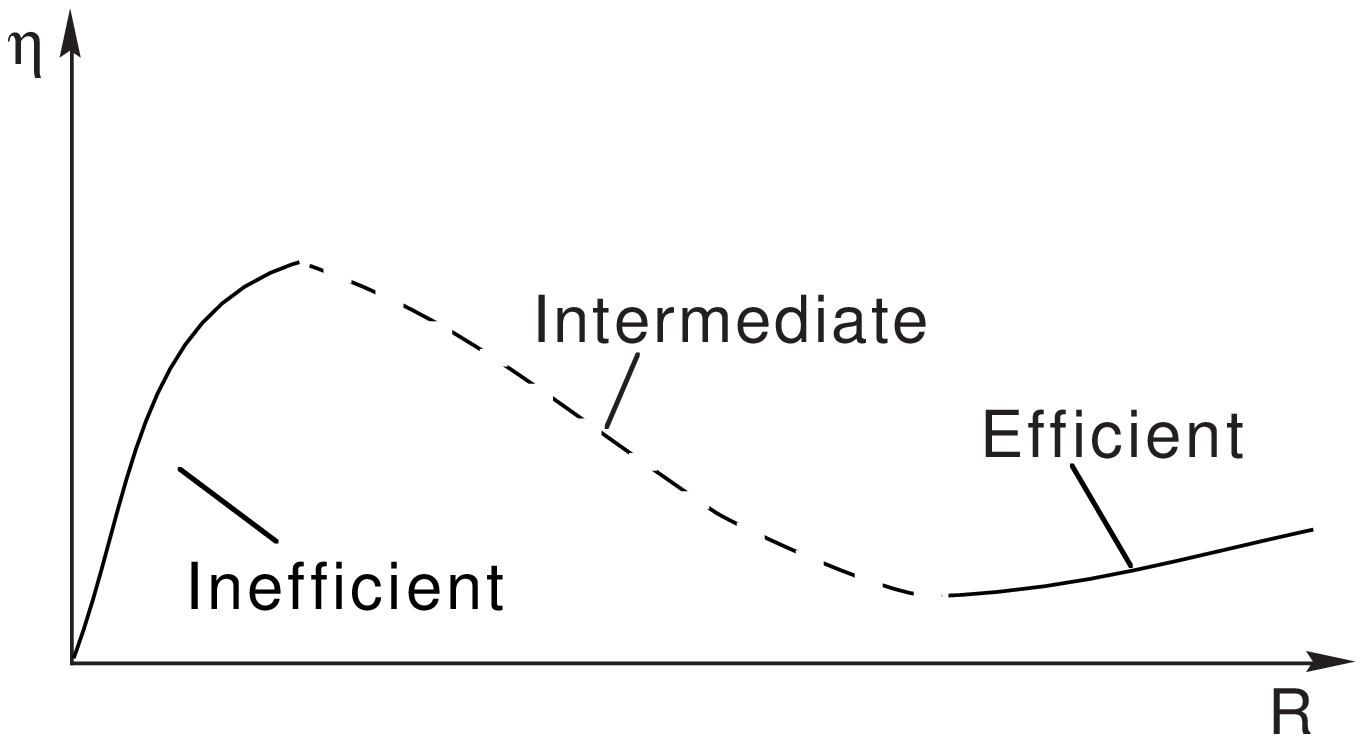]{Biffurcation curve for three possible
	solutions in the ($R,\eta $) plane. Here $R $ is the flow
	compression upstream, between the infinity and the subshock,
	$R=u_1/u_0 $ and $\eta $ is the injection rate. \label{fig1}}

	\figcaption[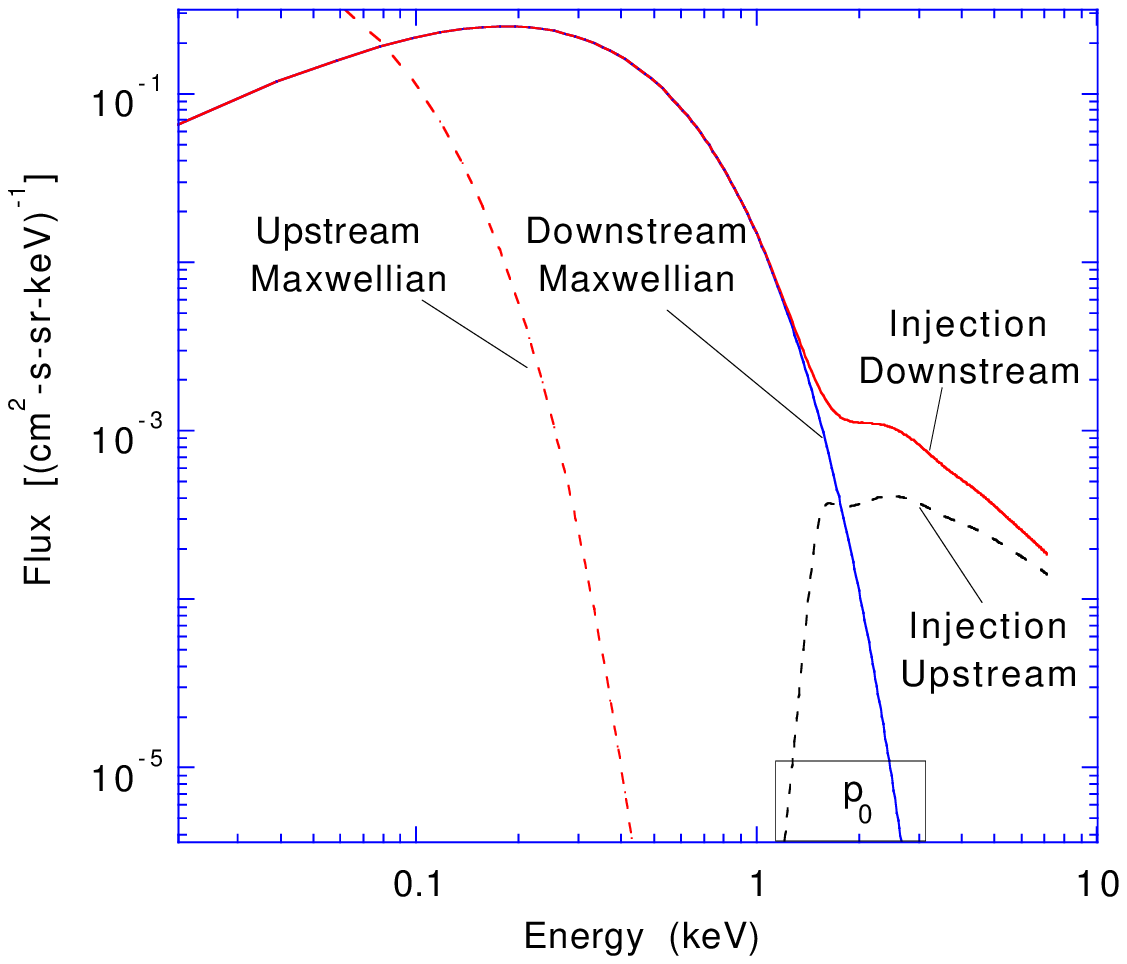]{Omnidirectional injection spectrum downstream
	(measured in the shock frame) calculated for the downstream temperature
	$ T_2=2\cdot 10^6 K $ and compression ratio $r_{\rm s}\approx 4$
	(MV97).  Dashed lines show the correspondent upstream distribution
	schematically.  A possible region for matching is indicated by the
	box.\label{fig2}}

	\figcaption[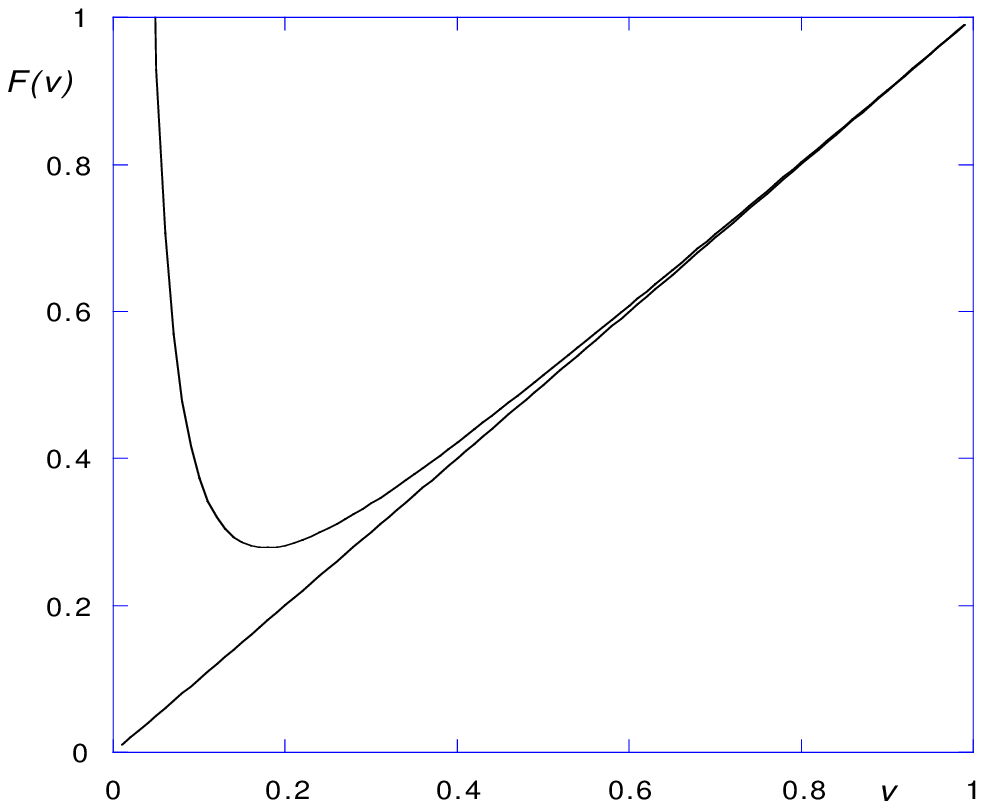]{The l.h.s of eq.(\ref{ber:v}) as a
	function of \( v \) for \( M = 10 \). \label{fig3}}

\end{document}